\documentclass[aps,prl,twocolumn,groupedaddress,showpacs,superscriptaddress,amssymb,amsmath]{revtex4-2}

\usepackage{graphicx}
\usepackage{dcolumn}
\usepackage{bm}
\usepackage{hyperref}
\hypersetup{
    colorlinks=true,
    linkcolor=blue,
    filecolor=magenta,      
    urlcolor=blue,
    pdftitle={Overleaf Example},
    pdfpagemode=FullScreen,
    }
\usepackage[mathlines]{lineno}
\usepackage{amsmath}
\usepackage [english]{babel}
\usepackage [autostyle, english = american]{csquotes}
\MakeOuterQuote{"}
\usepackage{float}
\begin{document}

\title{Tensile strain induced brightening of momentum forbidden dark exciton in WS$_2$}

\author{Tamaghna \surname{Chowdhury}} \email{tamaghna.chowdhury@students.iiserpune.ac.in}
\affiliation{Department of Physics, Indian Institute of Science Education and Research (IISER), Pune 411008, India.}
\affiliation{Department of Physics and Astronomy, University of Manchester, United Kingdom, M13 9PL.}
\author{Sagnik \surname{Chatterjee}}
\affiliation{Department of Physics, Indian Institute of Science Education and Research (IISER), Pune 411008, India.}
\author{Dibyasankar \surname{Das}}
\affiliation{Department of Condensed Matter Physics and Materials Science, Tata Institute of Fundamental Research,
Mumbai 400005, India.}
\author{Ivan \surname{Timokhin}}
\affiliation{Department of Physics and Astronomy, University of Manchester, United Kingdom, M13 9PL.}
\author{Pablo Díaz  \surname{Núñez}}
\affiliation{Department of Physics and Astronomy, University of Manchester, United Kingdom, M13 9PL.}
\author{Gokul \surname{M. A.}}
\affiliation{Department of Physics, Indian Institute of Science Education and Research (IISER), Pune 411008, India.}
\author{Suman \surname{Chatterjee}}
\affiliation{Department of Electrical Communication Engineering, Indian Institute of Science, Bangalore 560012, India}
\author{Kausik \surname{Majumdar}}
\affiliation{Department of Electrical Communication Engineering, Indian Institute of Science, Bangalore 560012, India}
\author{Prasenjit \surname{Ghosh}}
\affiliation{Department of Physics, Indian Institute of Science Education and Research (IISER), Pune 411008, India.}
\author{Artem \surname{Mishchenko}}
\affiliation{Department of Physics and Astronomy, University of Manchester, United Kingdom, M13 9PL.}
\author{Atikur \surname{Rahman}}
\email{atikur@iiserpune.ac.in}
\affiliation{Department of Physics, Indian Institute of Science Education and Research (IISER), Pune 411008, India.}


\begin{abstract}

Transition-metal dichalcogenides (TMDs) host tightly bound quasi-particles called excitons. Based on spin and momentum selection rules, these excitons can be either optically bright or dark. In tungsten-based TMDs, momentum-forbidden dark exciton is the energy ground state and therefore it strongly affect the emission properties. In this work, we brighten the momentum forbidden dark exciton by placing WS$_2$ on top of nanotextured substrates which put the WS$_2$ layer under tensile strain, modifying electronic bandstructure. This enables phonon assisted scattering of exciton between momentum valleys, thereby brightening momentum forbidden dark excitons. Our results will pave the way to design ultrasensitive strain sensing devices based on TMDs.
\end{abstract}
\maketitle



TMDs ({\it e.g.} $\mathrm{MX}_2$, $\mathrm{M}\!=\! \mathrm{Mo}$, W, $\mathrm{X}\!=\!\mathrm{S}$, Se) are known for their novel optical properties\cite{novoselov2005two,britnell2013strong,mak2010atomically,ross2013electrical}. They host excitons - charge neutral electron-hole pairs bound by Coulomb interactions.\cite{splendiani2010emerging, he2014tightly}. 
The large spin-orbit coupling in WS$_2$ due to the heavy mass of W atom, splits the valance band (VB) maxima and conduction band (CB) minima at $K$, $K'$ points in two sub-bands with opposite spin orientations (up, down at $K$ and down, up at $K'$) respectively. This results in the formation of two `bright' intravalley excitons with opposite spins at $KK$, $K'K'$\cite{malic2018dark}.  
There is possibility for the formation of indirect intervalley excitons $K\Lambda$ as well. But because of the large momentum mismatch, they require the assistance of phonons to recombine radiatively by emitting a photon\cite{brem2020phonon}.  The $K\Lambda$ exciton is therefore called momentum-forbidden dark exciton. In case of W-based ML TMDs, $K\Lambda$ exciton is the excitonic ground state and has higher binding energy and longer lifetime than the bright excitons $KK$, $K'K'$ and therefore they play an important role in the exciton dynamics of the system\cite{feierabend2019dark,madeo2020directly}. Thus, controlling them is essential for designing novel optical devices. 
 The dark exciton can be brightened by exciton-phonon coupling if the  energy of the available phonon mode matches with the dark-bright exciton energy splitting which can be effectively tuned by applying strain on ML WS$_2$\cite{selig2018dark,feierabend2019dark,zollner2019strain}.
Therefore, strain acts as a tuning knob for the emission of dark excitons\cite{niehues2018strain,feierabend2019dark}. 

\begin{figure}[t]
\includegraphics[width=\linewidth]{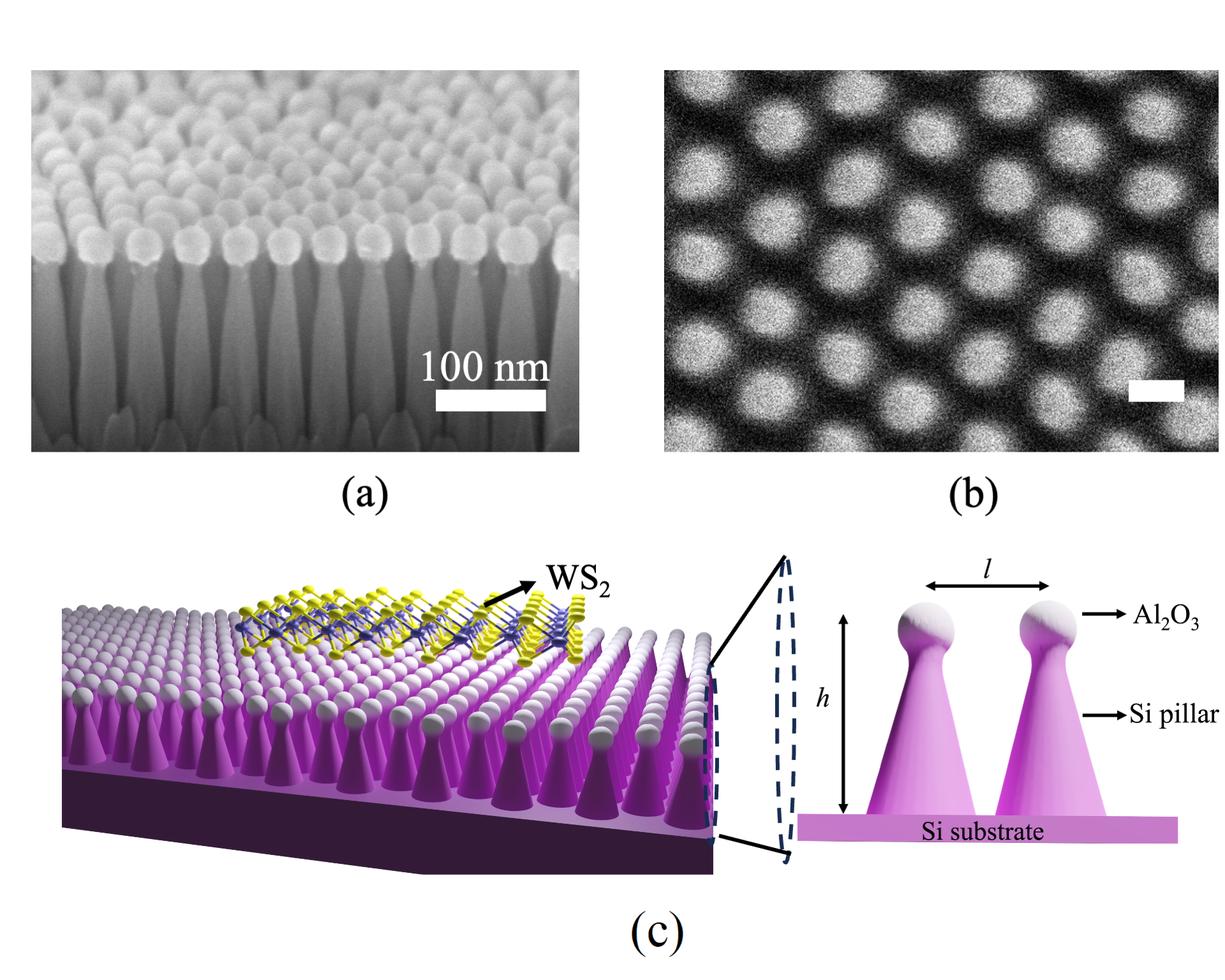}
\vspace*{-7mm}
\caption{\label{fig:1} (a) Field Emission Scanning Electron Microscope (FESEM) image of the C-99 nanotextured sample (sideview). (b) FESEM image of the C-99 substrate as viewed from top (scale bar in the inset is 20 nm). (c) Schematic of the substrate with ML WS$_2$ placed on top. The zoomed in view shows the details of various dimensions and constituent materials of the nanopillars.
}
\end{figure}

In this work, we apply tensile strain on ML WS$_2$ by placing them on nanotextured substrates patterened with nanopillars. The nanopillars of height `\textit{h}' and interpillar seperation (center to center) `\textit{l}' [Fig.\ref{fig:1}a, c]. Conically shaped nanopillars made of Si(100) have insulator Al$_2$O$_3$ nanospheres 10 nm on top[Fig.\ref{fig:1}c].
We prepared samples with varied interpillar distances: \textit{l} $\sim$ 25 nm(sample C-49), \textit{l} $\sim$ 44 nm (C-99) and \textit{l} $\sim$ 60nm (C-132). By tuning `\textit{l}' we tune the amount of strain applied on ML-WS$_2$. ML WS$_2$ was grown by chemical vapour deposition (CVD) and was transferred on top of the nanotextured substrate by wet transfer technique [Fig. \ref{fig:1}b] (details of sample preparation and characterization can be found in supplemental material section I, XI and our previous work\cite{chowdhury2021modulation, rahman2015sub}). We perform temperature dependent photoluminescence (PL) and Raman measurements on the strained and unstrained ML WS$_2$ samples. Supported by, \textit{ab initio} calculations, we discover the brightening of K$\Lambda$ dark excitons by applying tensile strain.  

\begin{figure}[t]
\includegraphics[width=\linewidth]{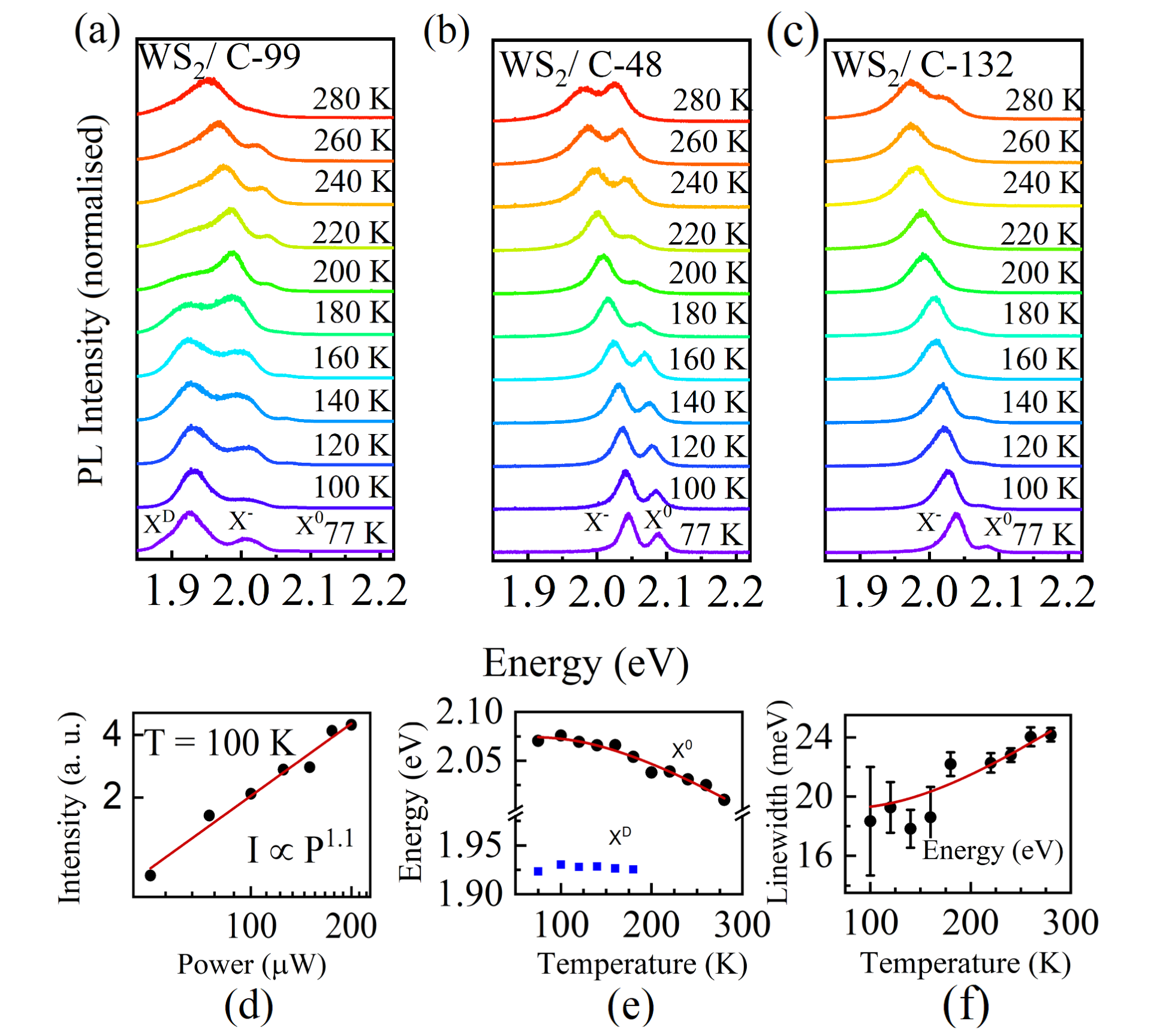}
\vspace*{-2mm}
\caption{\label{fig:2} Temperature dependent PL spectra of ML WS$_2$ placed on top of (a) C-99 (b) C-48 and (c) flat SiO$_2$/Si substrates. All the spectras are recorded at an excitation power of 50 $\mu$W. The spectras are shifted along y-axis for clarity. (d) The integrated intensity (I) of X$^D$ as function of excitation power (P) fitted with the relation I $\propto$ P$^\alpha$ where $\alpha$ is the exponent. (e) Temperature dependence of the peak position of X$^0$ (black solid circles) and X$^D$(blue solid squares), the temperature dependence of X$^0$ is fitted with Eq. \ref{eq:2} (red line). (f) FWHM of X$^0$ as a function of temperature fitted with Eq. \ref{eq:3}.
}
\end{figure}

 The PL measurements were performed using a continuous wave laser of wavelength 514.5 nm. The details of the measurement can be found in the supplemental material section II. In the temperature dependent PL study of ML WS$_2$ on C-99 substrate [Fig. \ref{fig:2}a], two well resolved peaks, one at $\sim$ 2.01 eV (FWHM$\sim$ 24 meV) and other at $\sim$1.95 eV (FWHM$\sim$ 32 meV) were observed at 280 K. The peak position and FWHM were extracted from each PL spectrum by fitting with a sum of Gaussian functions (see supplemental material section VI for details).
We attribute the peak at 2.01 eV as bright exciton $KK/K'K'$ peak (X$^0$) and the peak at 1.95 eV as negatively charged trion peak (X$^-$) since ML WS$_2$ is a n-doped semiconductor.We further confirm this attribution of peaks from excitation power (P) dependence of the integrated intensity (I). The data were fitted with the power law dependence I$\propto$ P$^\alpha$ and we obtained values for $\alpha$ of 0.9, 1.03 for X$^0$ and X$^-$ respectively, typical for excitons and trions (see supplemental material section VII)\cite{shang2015observation,paradisanos2017room,plechinger2015identification}. As we lower the temperature both X$^0$, X$^-$ blueshift as reported earlier\cite{plechinger2015identification}. At around 200 K a new peak starts to appear at $\sim$1.94 eV. As we further decrease the temperature, the intensity of new peak increases while the opposite is true for X$^-$ and X$^0$: their intensity diminish\cite{Feierabend2017}.  The new peak can be attributed to (a) a biexciton, XX (b) a defect bound exciton (X$^L$) or (c) a dark exciton (X$^D$). The exponent $\alpha$ of the power law dependence for XX and X$^L$ is known to be superlinear ($\sim$2.0) and sublinear ($\sim$0.5) respectively\cite{shang2015observation,paradisanos2017room,plechinger2015identification}. From the I vs P plot of the new peak we obtain a value of $\alpha$ of about $\sim$ 0.97 and 1.15 at 180 and 100 K respectively (see Fig. 2d and supplemental material section VII). Moreover, the new peak do not show any blueshift with increasing excitation power, characteristic of X$^L$ because of its broad energy distribution \cite{bluesfift_defect,saigal}. However, it showed red shift due to local heating, a behaviour generally seen in excitons \cite{chowdhury2021modulation} (see supplemental material section VIII). Furthermore, the new peak also shows anisotropy in circular polarization dependent PL, uncharacteristic of X$^L$ \cite{10.1063/1.4963133, saigal}(see supplemental information section IV for details). Therefore, the new peak is neither XX or X$^L$. However, value of $\alpha$ and its peak position at 77 K $\sim1.92$ eV is similar to recent reports of observation of dark exciton under strain and strong exciton-phonon coupling\cite{chand2022visualization}. We therefore attribute this new peak at $\sim1.92$ eV as X$^D$. We performed the similar study on other two samples namely C-48 and C-132 [Fig. \ref{fig:2}b and c]. For the C-48 and C-132 samples we observed X$^0$ and X$^-$ peaks at room temperature but no new peaks were observed as we lowered the temperature to 77 K. The X$^0$ and X$^-$ showed blueshift and narrowing with decreasing temperature similar to that of C-99 sample. 


The temperature dependence of the peak position of X$^0$ in C-99 sample was studied in detail [Fig. \ref{fig:2}e]. The temperature dependence shows a  characteristic redshift with decreasing temperature induced by a exciton-phonon coupling. This can be described by the phenomenological model proposed by O'Donnell and Chen\cite{o1991temperature}:
\begin{equation}
    E(T)= E(0)-S\langle\hbar\omega\rangle\Big(\coth{\frac{\langle\hbar\omega\rangle}{k_{\mathrm{B}}T}}-1\Big)
    \label{eq:2}
\end{equation}
where \textit{E(T)} is the resonance energy of X$^0$ at temperature \textit{T}, \textit{S} is dimensionless exciton-phonon coupling constant, \textit{$k_B$} is Boltzmann constant and $\langle\hbar\omega\rangle$ is the average phonon energy responsible for the coupling. By fitting the experimental data we obtained the parameters, E(0)= 2.074 $\pm$ 0.003 eV, S= 3.65 $\pm$ 0.98 and $\langle\hbar\omega\rangle$= 43 $\pm$ 10 meV. The value of $\langle\hbar\omega\rangle$ is close to the energy of E$^{'}$ phonon ($\sim$ 43.9 meV) mode of ML WS$_2$. This suggests that the E$^{'}$ phonon mode has a crucial role in the exciton-phonon coupling. Note that the peak position of X$^D$ changes only by $\sim$2 meV as we increase the temperature from 75 K to 180 K, whereas, in the same temperature range the X$^0$ peak position changes by $\sim$20 meV. This observation is consistent with the fact that the CB minima at K point shifts at a much faster rate with temperature compared to the $\Lambda$ point \cite{chand2022visualization}. To determine the strength of exciton-phonon coupling, the evolution of the FWHM of X$^0$ was fitted by a phonon-induced broadening model[Fig. \ref{fig:2}f]\cite{selig2016excitonic,cadiz2017excitonic}:
\begin{equation}
    \gamma=\gamma_{0}+c_1T+\frac{c_2}{e^{\frac{\hbar\omega}{k_{B}T}}-1}
    \label{eq:3}
\end{equation}
where $\gamma_0$ is the intrinsic FWHM, the linear term in T is due to the interaction of acoustic phonon modes (LA and TA) and the last term is the interaction term with the optical phonon mode\cite{rudin1990temperature}. c$_2$ is the measure of the exciton- optical phonon coupling strength. The value of $\hbar\omega$ that we obtained previously by fitting Eq. \ref{eq:2}, was used for fitting Eq. \ref{eq:3}. The value of c$
_2$ obtained by fitting Eq. \ref{eq:3} is 26.5 $\pm$ 4.6 and is significantly higher than the previously reported value of 6.5 for ML WS$_2$\cite{cadiz2017excitonic}. This higher value of c$_2$ further confirms the strong exciton and E$^{'}$ phonon mode coupling in C-99 substrate. See supplemental material section XII. for the above analysis of X$^D$ peak in ML WS$_2$ on top of C-48 and C-132 substrate.

We further did temperature dependent Raman study on the C-99 sample. Phonon modes responsible for electron-phonon scattering in case of ML WS$_2$ are LA, TA, E$^{'}$ and A$_1$ modes\cite{jin2014intrinsic}. The various Raman peaks (E$^{'}$, 2LA, A$_{1}^{'}$) were analysed with multiple Lorentzian functions[ Fig. \ref{fig:3}a]\cite{berkdemir2013identification} (see supplemental material section IX for fitting details).
All the phonon modes except the in-plane  E$^{'}$ mode showed redshift in Raman shift and an increase in their linewidth with increasing temperature [see Fig. \ref{fig:3}b, c for E$^{'}$ and supplemental material section X for A$_1^{'}$]. The redshift and increasing linewidth with temperature can be explained by anharmonic cubic equations\cite{balkanski1983anharmonic,joshi2016phonon}:
\begin{figure}[t]
\includegraphics[width=\linewidth]{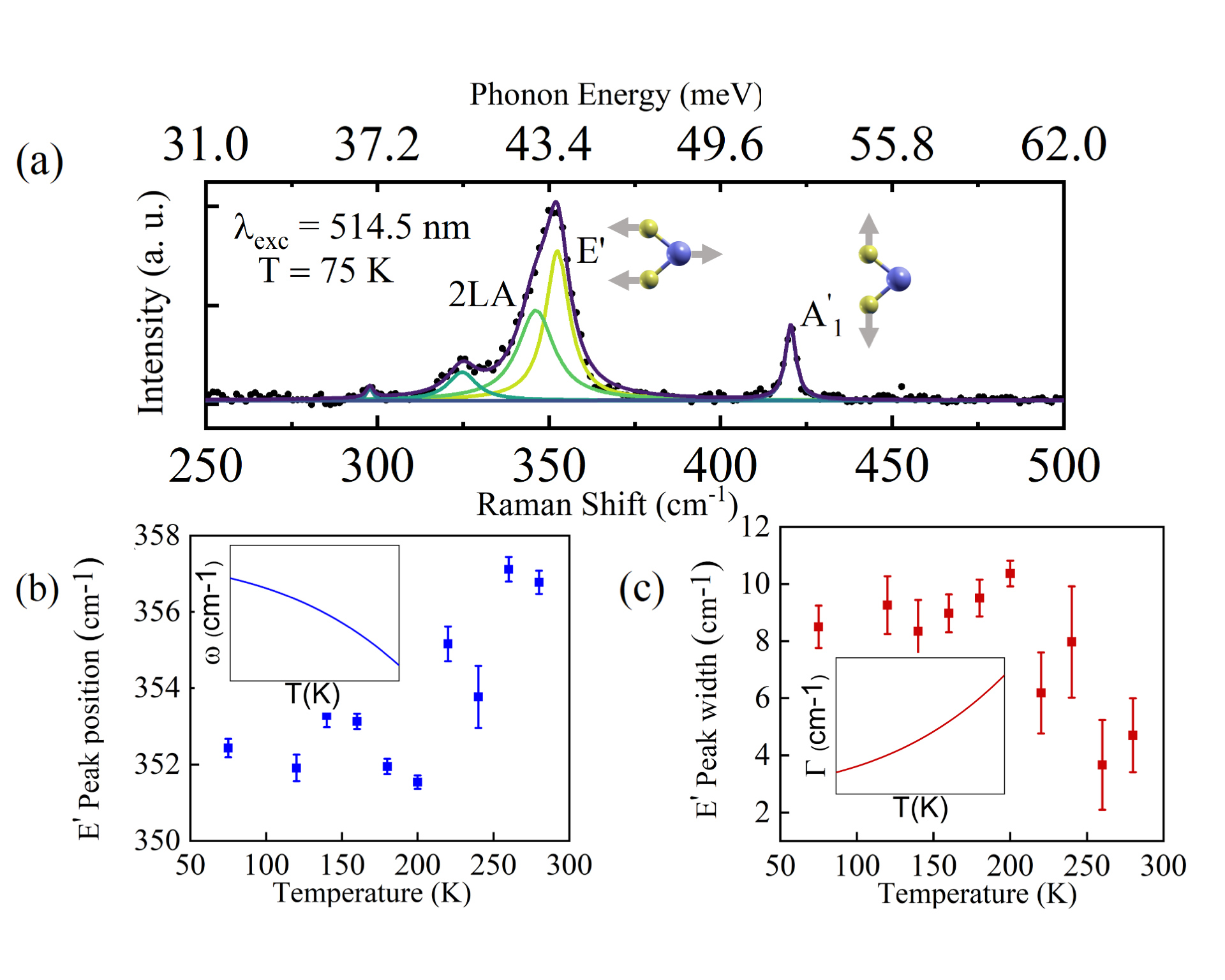}
\vspace{-7mm}
\caption{\label{fig:3} (a) Raman spectra of ML WS$_2$ placed on top of C-99 substrate at 75 K. Various peaks are labeled according to ref. \cite{berkdemir2013identification}. The main vibrational modes E$^{'}$ and A$_{1}^{'}$ are shown in the schematic. The blue and yellow balls represent tungsten (W) and sulphur (S) atoms respectively.  Anomalous behaviour of E$^{'}$ peak 
(b) position and (c) width as a function of temperature. Anharmonicity of the phonon mode as per Eq. \ref{eq:3} and Eq. \ref{eq:4} are in the respective insets.}
\end{figure}
\begin{equation}
    \omega_{ph}(T)= \omega_0 - C\Big(1+\frac{2}{e^\frac{\hbar\omega_0}{2k_BT}-1}\Big)
    \label{eq:4}
\end{equation}
\begin{equation}
    \gamma_{ph}(T)= \gamma_0 + D\Big(1+\frac{2}{e^\frac{\hbar\omega_0}{2k_BT}-1}\Big)
    \label{eq:5}
\end{equation}

where $\omega_{ph}(T)$, $\gamma_{ph}(T)$ are the frequency of the phonon mode and linewidth at temperature T respectively. $\omega_0$, $\gamma_{0}(T)$ are the frequency of phonon mode and linewidth at T= 0 K respectively and \textit{C} is a constant. The behaviour of Eq. \ref{eq:4} and Eq. \ref{eq:5} as a function of temperature is plotted in the insets of Fig. \ref{fig:3}b and c. The E$^{'}$ phonon modes shows a completely opposite trend when compared to Eq. \ref{eq:4} and Eq. \ref{eq:5} [Fig. \ref{fig:3}b, c]. This anomalous behaviour of E$^{'}$  phonon mode is related to strong electron-phonon coupling\cite{ferrari2007raman,bonini2007phonon,chae2010hot}. 
The various factors affecting the Raman mode frequency can be expressed mathematically as \cite{paul2020tailoring} $\omega(T) =\omega_0+\Delta\omega_{vol}(T)+\Delta\omega_{anh}(T)+\Delta\omega_{sp-ph}(T)+\Delta\omega_{e-ph}(T)$ where $\Delta\omega_{vol}(T)$ corresponds to quasiharmonic contribution due to change in unit-cell volume, $\Delta\omega_{anh}(T)$ corresponds to phonon-phonon interaction related anharmonic effects, $\Delta\omega_{sp-ph}(T)$ is due to spin-phonon coupling and $\Delta\omega_{e-ph}(T)$ is due to electron-phonon coupling. Eq. \ref{eq:4} and \ref{eq:5} take into account the first three terms but not the last term. Therefore Eq. \ref{eq:4} and \ref{eq:5} fails to describe the anomalous behaviour of the E$^{'}$ Raman modes. 

To understand the effect of strain, calculated the electronic band structure of ML-WS$_2$ by DFT using full-relativistic ultrasoft pseudopotential with Perdew–Burke–Ernzerhof (PBE) exchange-correlation functionals alongside plane waves implemented in Quantum ESPRESSO package [Fig.\ref{fig:4}a] (see supplemental material section V for details). With increasing tensile(compressive) strain the absolute CB minima at K- point shifts down (up) and the local CB minima at $\Lambda$ shifts up (down). The VB maxima at K and $\Lambda$ shows almost no change with strain[Fig. \ref{fig:4}a]. We denote the direct bandgap at K point as E$^{KK}$ and the indirect bandgap at $\Lambda$ point as E$^{K \Lambda}$. Note that, in electronic band structure the CB minima at $\Lambda$ point is at higher energy  than the CB minima at K point by E$^{K\Lambda}$ - E$^{KK}$ = $\Delta E^{K\Lambda}$= 64 meV. To get into the exciton picture from electron-hole picture (as described in ref. \cite{feierabend2019dark,malic2018dark}) we need to calculate the binding energy of the excitons. The binding energy (E$_b$) is calculated from the effective mass model \cite{chernikov2014exciton}:
\begin{equation}
    E_b= \frac{\mu e^4}{2\hbar^2\epsilon^2(n-\frac{1}{2})^2}
\label{eq:7}
\end{equation}

where $\frac{1}{\mu}\!=\!\frac{1}{m_e}+\frac{1}{m_h}$ the exciton reduced mass, m$_h$ and m$_e$ are the effective masses of holes and electrons respectively (see supplemental material section XIV for details). $\epsilon$ is the dielectric constant of ML WS$_2$, \textit{e} is the electron charge and n is the principal quantum number. m$_e$ and m$_h$ are calculated from parabolic band approximation of the electronic bands obtained from DFT calculations (extracted values are listed in Table \ref{tab:1})\cite{ashcroft2021solid}. $\epsilon$ of ML-WS$_2$ was taken to be $\sim$ 5 for n$=1$ as was shown in \cite{chernikov2014exciton}.  $\mu$ of K-K exciton (X$^0$) and K-$\Lambda$ exciton (X$^D$) are $\sim$  0.155m$_0$ and 0.219m$_0$ respectively, where m$_0$ is the free electron mass. Using Eq. \ref{eq:7} we found E$_b$ of X$^0$ and X$^D$ $\sim$ 310 meV and 438 meV respectively. Now, if we visualize the scenario in the excitonic picture, the K-K bright exciton state is formed at the $\Gamma$ point (zero momentum point) in the center of mass (COM) coordinates at the position E$^{KK}_{exc}$ = E$^{KK}$ - E$_{b}^{X^0}$ and the K-$\Lambda$ dark exciton state is formed at the $\Lambda$ point in COM coordinates at the position E$^{K\Lambda}_{exc}$ = E$^{K\Lambda}$ - E$_{b}^{X^D}$\cite{Bao_2020}. As the E$_{b}^{X^D}$ is higher than E$_{b}^{X^0}$, X$^D$ is at a lower energy than X$^0$ in the exciton picture, unlike the electron-hole picture. Our calculations show that, in unstrained sample, the dark state is below the bright state by energy E$^{KK}_{exc}$ - E$^{K\Lambda}_{exc}$ = $\Delta$E $\sim$ $64$ meV. Under strain, the excitonic states at $\Gamma$ (COM) and $\Lambda$ (COM) points behave similar to the CB minima at K and $\Lambda$ point respectively. The change of $\Delta$E as a function of strain is plotted in Fig. \ref{fig:4}c (the values used to generate this plot can be found in supplementary section V). The fitting shows $\sim$ 184 $\pm$ 2.5 meV change of $\Delta$E with 1\% of applied strain. Note that E$_b$ does not change that much with strain as the latter has little influence on $\mu$\cite{PhysRevB.96.045425, feierabend2019dark}.
\begin{table}
    \centering
\caption{The values of effective mass of electron at the CB minima of K point, $\Lambda$
point (m$_{K}^{e}$ and m$_{\Lambda}^{e}$ respectively) and VB maxima at K point (m$_{K}^{h}$).}
\label{tab:1}
    \begin{tabular}{|c|c|c|c|c|} \hline 
         m$_{K}^{e}$&  m$_{K}^{h}$&  m$_{\Lambda}^{e}$&  $\mu_{K-K}$& $\mu_{K-\Lambda}$\\ \hline 
         0.272 m$_0$&  0.358 m$_0$&  0.563 m$_0$&  0.155 m$_0$& 0.219 m$_0$\\ \hline
    \end{tabular}

\end{table}

\begin{figure}[t]
\includegraphics[width=\linewidth]{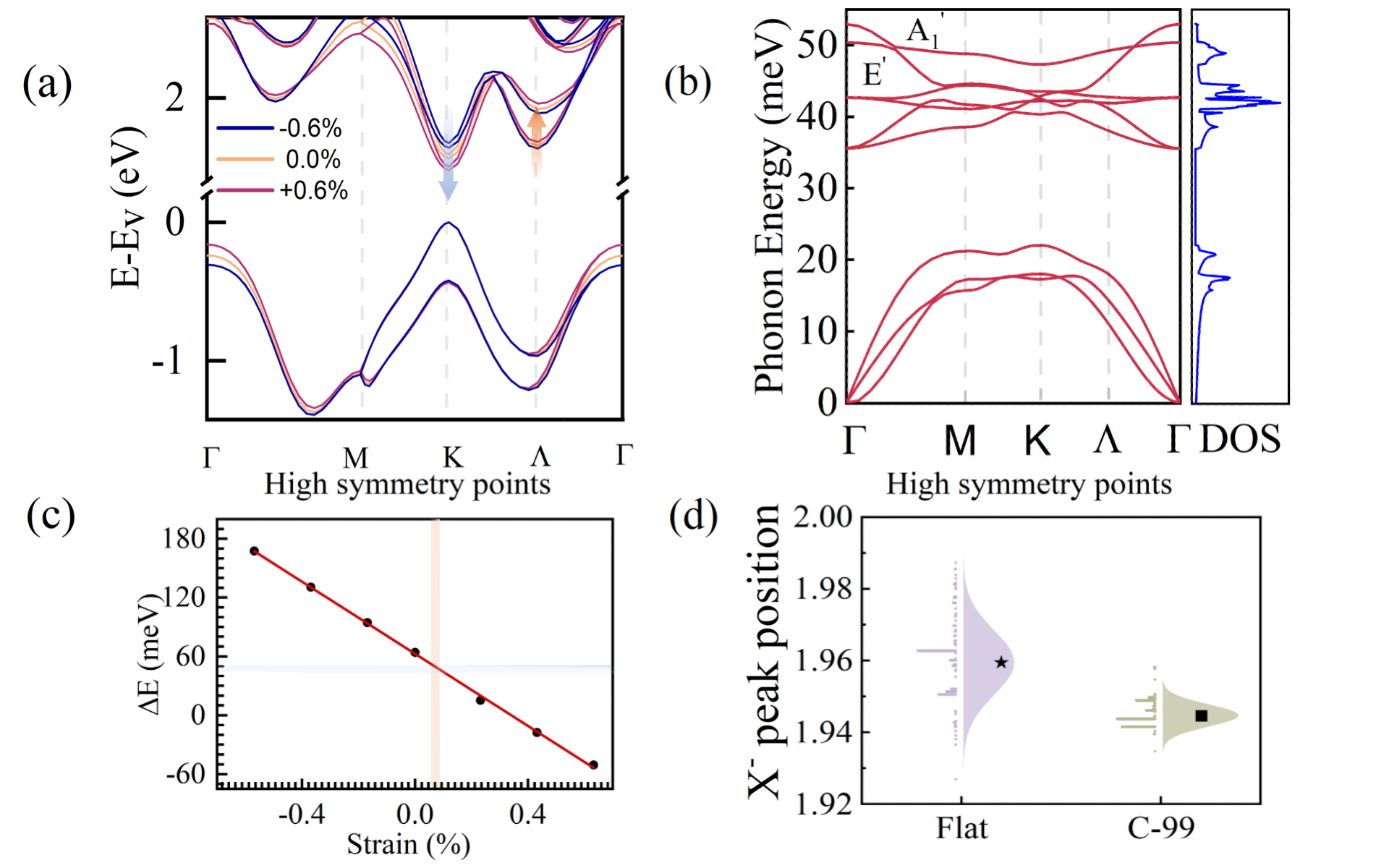}
\vspace*{-1mm}
\caption{\label{fig:4}(a) DFT calculated electronic band diagram (considering spin-orbit interaction) of ML WS$_2$ with different strain values -0.6\%, 0\% and +0.6\% respectively. To focus on relative changes, VB maxima at the K point is used a reference. 
(b) Phonon dispersion of ML WS$_2$ (left panel), phonon density of states (DOS) (right panel). (c)
$\Delta$E as a function of strain on ML WS$_{2}$. The data was fitted with a straight line (red). 
 The orange shaded area lies in the region of E$^{'}$ phonon mode and yellow shaded area shows the corresponding range of strain.
(d) Mean peak position of X$^-$ in ML WS$_2$ on top of SiO$_2$/Si (flat) and C-99 extracted from the PL map. The normal distribution and the raw data points are also shown.}

\end{figure}

\begin{figure}[t]
\includegraphics[width=\linewidth]{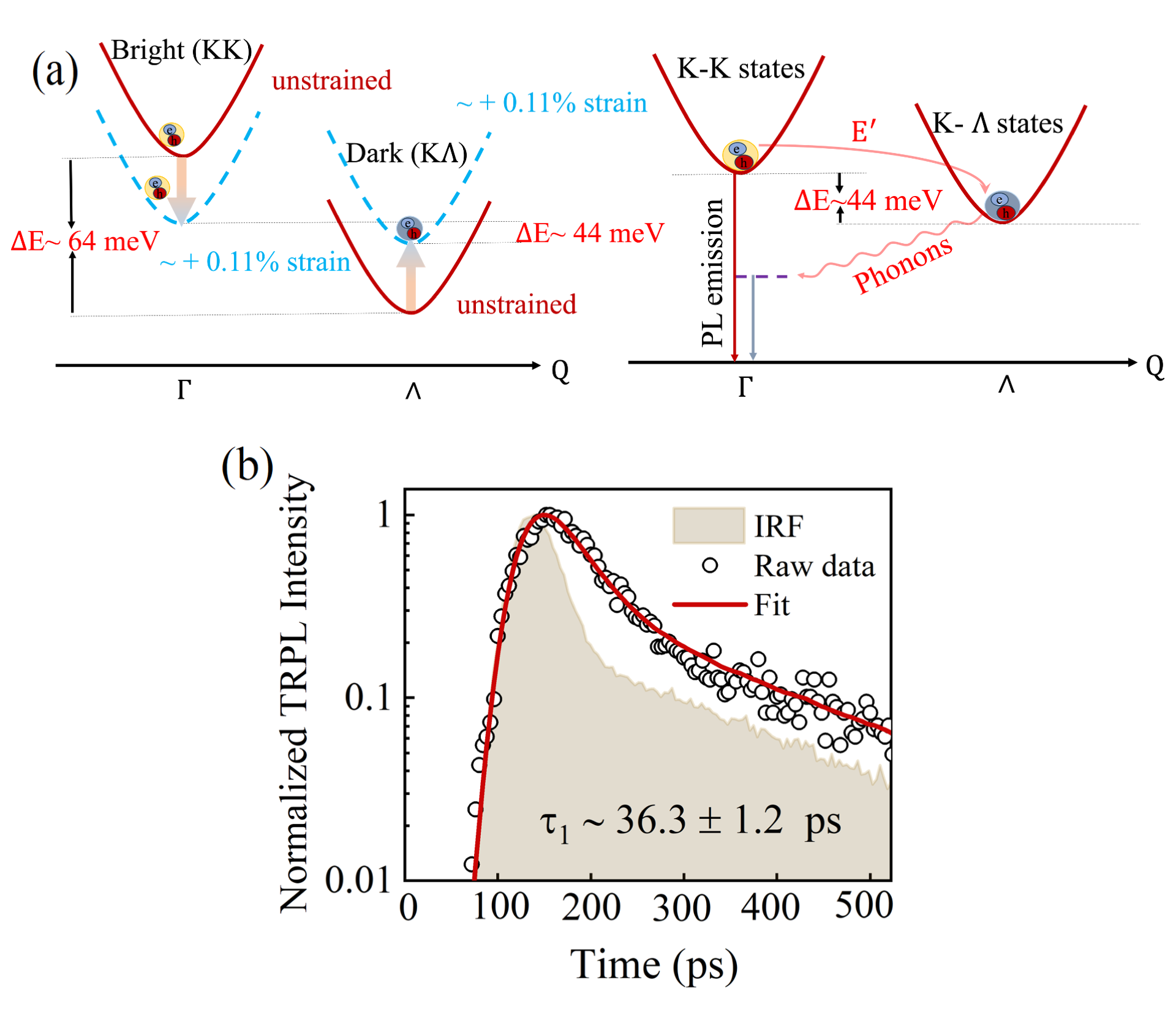}
\vspace*{-3mm}
\caption{\label{fig:5} (a) Schematic showing the E$^{'}$ phonon mediated scattering of the bright KK X$^0$ excitons from the $\Gamma$ point (COM) to $\Lambda$ point (COM) due to tensile strain. 
(b) TRPL measurement on ML WS$_2$ at 60 K to determine the lifetime of X$^D$ peak. 
}
\end{figure}

 On optically exciting a coherent exciton population at $\Gamma$ point (COM), K-K bright excitons are formed. Incoherent excitons are then formed at $\Lambda$ point (COM) by phonon assisted scattering of excitons from $\Gamma$ point (COM), where a phonon covers the energy and momentum mismatch\cite{selig2018dark}. However, in the unstrained case, no optical or acoustic phonon modes with energy  $\Delta$E $\sim$ 64 meV are available, therefore K-$\Lambda$ states are not formed at $\Lambda$ point (COM). Whenever we apply tensile strain on the ML WS$_2$, $\Delta$E decreases and under $\sim$0.11 $\pm$ 0.01\% strain the value of $\Delta$E is $\sim$ 44 meV. From the PL map (see supplemental material section XIII) of X$^-$ and X$^0$ the distribution of their positions was plotted. The statistical distribution was fitted with a normal distribution to extract the mean and standard deviation [Fig. \ref{fig:4}d]. To estimate the amount of strain on ML WS$_2$ on top of C-99 substrate due to nanopillars, its position of X$^-$ was compared with the X$^-$ position in ML WS$_2$ on top of flat SiO$_2$/Si [Fig. \ref{fig:4}d]. ML WS$_2$ on top of flat SiO$_2$/Si  was considered to be unstrained. We did not take into account X$^0$ position for this purpose because X$^0$ was not clearly resolved in ML WS$_2$ on top of flat SiO$_2$/Si. The mean X$^-$ position was found to be 1.96 eV and 1.94 eV for SiO$_2$/Si and C-99 respectively. This amounts to $\sim$ 20 meV redshift of X$^-$ in C-99. It is reported that X$^-$ and X$^0$ redhshifts by $\sim$ 130 and 127 meV respectively for 1\% applied tensile strain\cite{michail2023tuning}. Therefore we can estimate that ML WS$_2$ on top of C-99 is under a tensile strain of $\sim$ 0.15 \%. 
 In C-99 sample phonon assisted scattering of excitons from $\Gamma$ point (COM) to $\Lambda$ valley (COM) is possible, thereby forming a population of K-$\Lambda$ excitonic states in the $\Lambda$ point (COM). The scattering process is illustrated in the schematic Fig. \ref{fig:5}a. An E$^{'}$ phonon with momentum $\Lambda$ (since at $\Gamma$ point (COM) momentum is zero) and energy $\sim$44 meV can make this scattering possible. The calculated phonon density of states shows a large number of phonon states available at $\sim$44 meV, thereby making this scattering more favorable [Fig. \ref{fig:4}b]. Note that, change of phonon energy with strain is very negligible\cite{Khatibi_2019}. 
The K-$\Lambda$ states at $\Lambda$ point (COM) then can scatter non-radiatively to a virtual state inside the light cone at $\Gamma$ point (COM) by emitting phonons. Once inside the light cone, the `dark' excitons can decay radiatively from the virtual state by emitting photon, thus leaving its signature in the PL spectra. 

To know about the kinetics of X$^D$, we did time-resolved PL (TRPL) on the ML WS$_2$ on top of C-99 substrate (see measurement details in supplemental material section III).
The measured TRPL data was fitted with two exponentials ($\sum_{n=1}^{2} A_{i}e^\frac{-t}{\tau_i}$) after deconvoluting from the IRF as implemented in QuCoa software (PicoQuant)[Fig.\ref{fig:5}b]. The faster and stronger component $\tau_1$ which represents the X$^D$ decay time is estimated to be $\tau_1 \sim$ 36.3 $\pm$ 1.2 ps. This value of $\tau_1$ is $\sim$ 30 times larger than the reported decay time of a neutral exciton X$^0$ ( $\tau$ $\sim$ 1 ps at T = 60 K) in literature \cite{TRPL1, TRPL2} . This longer lifetime of X$^D$ compared to X$^0$ is expected because, X$^D$ is excitonic ground state of ML WS$_2$\cite{chand2022visualization, madeo2020directly}. The slower ($\frac{A_1}{A_2}\sim 294$) and weak decay component $\tau_2 \sim $ 100 ps is expected to be coming from the contribution of tail of defect-bound exciton complex observed in ML WS$_2$ at lower temperatures\cite{Chatterjee_2022}.

In summary, we have reported the experimental observation of momentum-forbidden K-$\Lambda$ dark excitons by applying tensile strain on ML WS$_2$ using a nanotextured substrate. 
The 2D TMDs are known to buckle easily with compressive strain\cite{duerloo2014structural} and is also more difficult to create, especially at low temperatures which is essential to prevent thermally activated depopulation of dark state into the bright state. However, it's easy to create tensile strain in the 2D TMDs and they can endure high values of tensile strength as well
. Therefore it would be more practical and application-oriented if we can modulate the dark exciton with tensile strain rather than compressive strain.



\bibliographystyle{apsrev4-2}
\bibliography{refs}
\textit{Acknowledgements}. The research reported here was funded by the Commonwealth Scholarship Commission and the Foreign, Commonwealth and Development Office in the UK (Grant no. INCN-2021-049). T. C. is grateful for their support. All views expressed here are those of the author(s) not the funding body.
 T. C. thanks the Prime Minister's Research Fellowship (PMRF), Government of India (ID: 0700441) for funding. AR acknowledges funding support from DST SERB Grant no. CRG/2021/005659 and partial funding support under the Indo-French Centre for the Promotion of Advanced Research (CEFIPRA), project no. 6104-2. We thank National Supercomputing Mission (NSM) for providing computing resources of `PARAM Bramha' at IISER Pune, which is implemented by C-DAC and supported by the Ministry of Electronics and Information Technology (MeitY) and Department of Science and Technology (DST), Government of India. The authors would like to acknowledge funding from National Mission on Interdisciplinary Cyber-Physical Systems (NM-ICPS) of the Department of Science and Technology, Govt. Of India through the I-HUB Quantum Technology Foundation, Pune, India. We acknowledge Professor Sandip Ghosh, TIFR Mumbai, India for his help in polarization PL experiments and valuable discussions.



\end{document}


\author{Tamaghna \surname{Chowdhury}}
\email{tamaghna.chowdhury@students.iiserpune.ac.in}
\affiliation{Department of Physics, Indian Institute of Science Education and Research (IISER), Pune 411008, India.}
\affiliation{Department of Physics and Astronomy, University of Manchester, United Kingdom, M13 9PL.}
\author{Sagnik \surname{Chatterjee}}
\affiliation{Department of Physics, Indian Institute of Science Education and Research (IISER), Pune 411008, India.}
\author{Dibyasankar \surname{Das}}
\affiliation{Department of Condensed Matter Physics and Materials Science, Tata Institute of Fundamental Research,
Mumbai 400005, India.}
\author{Ivan \surname{Timokhin}}
\affiliation{Department of Physics and Astronomy, University of Manchester, United Kingdom, M13 9PL.}
\author{Pablo Díaz  \surname{Núñez}}
\affiliation{Department of Physics and Astronomy, University of Manchester, United Kingdom, M13 9PL.}
\author{Gokul \surname{M. A.}}
\affiliation{Department of Physics, Indian Institute of Science Education and Research (IISER), Pune 411008, India.}
\author{Suman \surname{Chatterjee}}
\affiliation{Department of Electrical Communication Engineering, Indian Institute of Science, Bangalore 560012, India}
\author{Kausik \surname{Majumdar}}
\affiliation{Department of Electrical Communication Engineering, Indian Institute of Science, Bangalore 560012, India}
\author{Prasenjit \surname{Ghosh}}
\affiliation{Department of Physics, Indian Institute of Science Education and Research (IISER), Pune 411008, India.}
\author{Artem \surname{Mishchenko}}
\affiliation{Department of Physics and Astronomy, University of Manchester, United Kingdom, M13 9PL.}
\author{Atikur \surname{Rahman}}
\email{atikur@iiserpune.ac.in}
\affiliation{Department of Physics, Indian Institute of Science Education and Research (IISER), Pune 411008, India.}

\title{Supplemental Material: Tensile strain induced brightening of momentum forbidden dark exciton in WS$_2$}
\maketitle

\section{Sample preparation}
The NT silicon samples were prepared by using block-copolymer lithography and inductively-coupled plasma etching (ICP-RIE). Please see our our earlier publication for the details of the NT substrate preparation. The NT substrates are patterned with nanopillars of interpillar distance `l'. The nanopillars made of Si(100) have cone like shape with insulator Al$_2$O$_3$ balls of diameter 10 nm on top of each nanopillar(see Figure ). We prepared sample of three different l= 25$\pm$3, 52$\pm$6 and 60$\pm$8 nm  by varying the molecular weight `M$_W$'= 48, 99 and 132 kg mol$^{-1}$ of the cylindrical phase polystyrene block-poly(methyl methacrylate) (PS-b-PMMA) block copolymer respectively. We name the three different substrates according to the molecular weight of PS-b-PMMA used for their preparation  as C-49(l= 25nm), C-99(l= 52 nm) and C-132(l= 60 nm). 
ML WS$_2$ were grown by APCVD on 300 nm thermally grown SiO$_2$ on Si. Properly cleaned and O$_2$ plasma-treated (60 W, 5 min) substrates were placed on an alumina boat containing 500 mg WO$_3$. A boat containing WO$_3$ powder was placed inside a 35 mm quartz tube in the heating zone of the furnace. Another boat containing 500 mg sulphur was placed upstream inside the tube, 15 cm away from the WO$_3$ boat. The sulphur boat was outside the heating zone and was heated using a heater coil. The tube was flushed with 500 sccm Ar for 20 min. The furnace was kept at 850 °C and the heater coil was kept at 240 °C for evaporation of sulphur. This set of temperatures was maintained for 10 min. After the growth, the system was allowed to cool naturally.
The ML WS$_2$ was then transferred from the growth substrate to the NT substrates and another  flat 300 nm SiO$_2$/ Si substrate by wet-transfer method using polystyrene (PS)\cite{chowdhury2021modulation, doi:10.1021/nn5057673}.
 \section{Temperature dependent PL and Raman measurements}

 The temperature dependent PL and Raman measurements were done using continuous flow Oxford Instruments optical cryostat HiRes. The PL and Raman spectra were obtained in a WiTec Alpha 300 spectrometer mounted with a grating of 600 and 1800 lines/mm respectively using a Nikon 100X objective (numerical aperture (N. A.) = 0.80) lens. The samples were excited with a continuous wave laser of wavelength 514.5 nm and spot diameter was $\sim$0.4 $\mu$m. A water-cooled charged couple device (CCD) was used for detection. All measurements were done at a low pressure of $2 X 10^{-5}$mbar pressure. The spectral resolution was $\sim$ 0.5 meV for PL and $\sim$ 1.3 cm$^{-1}$ for Raman measurements. Liquid nitrogen was used as cryogen and temperature was varied between 77 K and 300 K. All measurements were done at a excitation power less than 1 mW to prevent laser heating related damage to the sample. PL and Raman mappings were done at room temperature and ambient pressure by placing the sample on a piezostage using a Nikon objective lens of 100X magnification and 0.95 N. A.
  \section{Time resolved photolumniscence (TRPL) measurement}

 The sample was excited with a 531 nm pulsed laser having FWHM of 48 ps and a repetition rate of 5 MHz. The sync signal from the laser driver is fed to channel 0 of a TCSPC (PicoHarp 300). The emission from the sample is fed to a single photon detector (Micro Photon Devices), the output of which is connected to channel 1 of the TCSPC. The IRF shows an FWHM of 52 ps and a decay time scale of 23 ps. Using deconvolution, we can acxcurately estimate down to the 10\% of the IRF width. To filter out X$^D$ only, we use a bandpasss filter (FWHM of 10 nm) centered at 635 nm. For \textit{in situ} steady-state PL spectra, a 50:50 beam splitter is used to divert part of the emission to a spectrometer.
\section{Circular polarization dependent PL }
Micro-PL spectroscopy measurements, with a spatial resolution of $\sim$ 1.5 $\mu$m, were performed using circularly polarized 532 nm (2.33 eV) laser. The micro-PL setup has a 0.55 m focal length monochromator coupled to a microscope arrangement built using a long working distance (WD) objective (magnification 50x, numerical aperture 0.5, WD =12 mm). The sample was cooled using a liquid nitrogen cryostat with optical access. The PL signal was detected using a thermoelectric cooled Silicon charged coupled device. For circularly polarized PL excitation we used a polarizer and a quarte waveplate with the laser, and identical combination at the detection end. A 532 nm narrow band pass filter was used in front of laser and another 532 nm sharp long pass filter was used in front of the monochromator to block detection of the scattered laser.

\begin{figure}[H]
  \includegraphics[width=10cm]{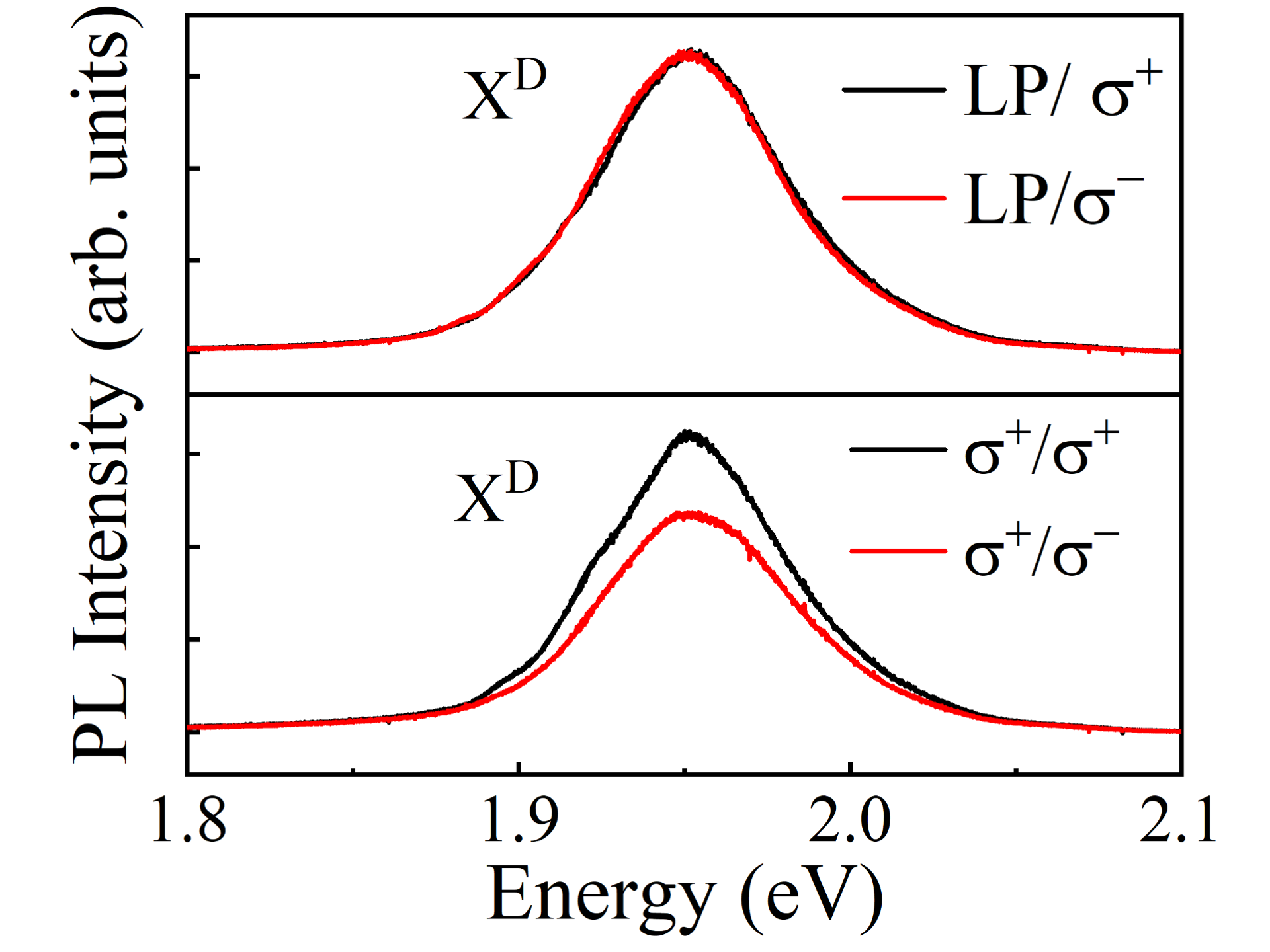}
  \centering
  \caption{PL spectra of the X$^D$ peak for $\sigma^{+}$  and $\sigma^{-}$ polarized
detection after excitation with linearly polarized light (upper panel) and $\sigma^{+}$ light (lower panel).
  \label{graph2}}
\end{figure}
Sample is excited with right-handed circularly polarised ($\sigma^{+}$) laser and the emitted PL is measured in both $\sigma^{+}$ and left-handed ($\sigma^{-}$) polarisation. The X$^D$ peak showed a degree of polarization (DoP)= $\frac{I_{\sigma^{+}}-I_{\sigma^{-}}}{I_{\sigma^{+}}+I_{\sigma^{-}}}$ of magnitude 13\%  even with a off-resonant excitation (off by $\sim$ 380 meV from X$^D$ exciton resonance energy), similar to earlier report\cite{TRPL1}.The I$_{\sigma^{+}/\sigma^{-}}$ was obtained by spectrally integrated Gaussian fit of the X$^D$ peak (Fig. \ref{graph2} lower panel)\cite{TRPL1}. To confirm that the weak polarization anisotropy of X$^D$ is indeed due to valley effects and not measurement artefacts, we excited the sample with linearly polarized light which will initially populate the K and K$'$ valley equally. The emitted $\sigma^{+}$ and $\sigma^{-}$ light showed $\sim$ 0 \% DoP(Fig. \ref{graph2} upper panel). This confirms that the polarization anisotropy is due to coupling of spin and valley degrees of freedom and origin of X$^D$ is not related to localization of excitons at the defect site (X$^L$)\cite{10.1063/1.4963133, saigal}.

 \section{DFT Calculations}
First-principle
density functional theory based calculations have be performed using
the Quantum ESPRESSO suite of codes \cite{Giannozzi_2009, Giannozzi_2017}. The electron-ion
interactions were described by ultrasoft pseudopotentials \cite{vanderbilt}.
TMDs show strong spin-orbit interactions which needs to be
incorporated in calculation of its electronic structure \cite{spin-orbit} . To
correctly account for this, we have solved the relativistic Kohn-Sham
equations self-consistently \cite{Kohn_Sham_1,Kohn_Sham_2}
using fully relativistic
ultrasoft pseudopotentials for all elements \cite{ultrasoft}. The
electron-electron exchange correlation functional have been described
using the Perdew-Bruke-Ernzerhof (PBE) parametrization of generalized
gradient approximation (GGA) \cite{PhysRevLett.77.3865}. The Brillouin zone (BZ) was
sampled with 6×6×1 Monkhorst-Pack k-point grid for BZ integrations. The
wave function energy cut-off is set to be 60 Ry amd the energy
cut-off for charge density is set to be 550 Ry. The optimized lattice
constant is obtained to be a$_0$= 3.18 \AA \! for unstrained case in good agreement with previous studies\cite{prevstudy}. The strained ML WS$_2$ was modeled via change of the lattice constant relative to strain with 
\begin{equation}
  a_0^s= a_0(1+\epsilon_s) 
  \label{eq:1}
\end{equation}

where a$_0^s$ is strained lattice constant,  a$_0$ is unstrained lattice constant and $\epsilon_s$ is applied strain.
We relax the new strained lattice with the lattice vectors modeled by the Eq (\ref{eq:1}).

The phonon dispersion and density of states (DoS) of ML WS$_2$ is calculated using the density functional perturbation theory (DFPT) method within the phonon package of Quantum ESPRESSO\cite{RevModPhys.73.515}. We sampled the momentum space with an 6×6×1 Monkhorst - Pack q-point mesh grid. Although, the change in the electronic band structure due to strain is much pronounced, the phonon dispersion does not change considerably with strain. The change in phononic dispersion is less than 0.4 \% per percent strain \cite{Khatibi_2019}. Therefore their energy and momentum can be considered constant in the range of strain in discussion in this paper.
The phononic dispersion shows that E$^{'}$ phonon of momentum $\Lambda$ required for the scattering of bright KK exciton to the $\Lambda$ valley to form dark K$\Lambda$ exciton is available with considerable density of states.

\begin{table}
    \centering
\caption{Values used for generating Fig. 4c of main text}
\label{tab:my_label}
    \begin{tabular}{|c|c|c|c|c|c|}\hline 
         Strain     ($\%$)&  E$^{KK}_{el}$(eV)&  E$^{K\Lambda}_{el}$(eV)&  E$^{KK}_{ex}=$E$^{KK}_{el}-$E$_{b}^{X^{0}}$ (eV)&  E$^{K\Lambda}_{ex}=$E$^{K\Lambda}_{el}-$E$_{b}^{X^{D}}$ (eV)& E$^{KK}_{ex}-$E$^{K\Lambda}_{ex}$ (meV)\\ \hline 
         -0.5686&  1.6735&  1.634&  1.3635&  1.196& 167.5\\ \hline 
         -0.3685&  1.6444&  1.6417&  1.3344&  1.2037& 130.7\\ \hline 
         -0.1684&  1.6156&  1.6492&  1.3056&  1.2112& 94.4\\ \hline 
         0&  1.5917&  1.6556&  1.2817&  1.2176& 64.1\\ \hline 
         0.2317&  1.5563&  1.6691&  1.2463&  1.2311& 15.2\\ \hline 
         0.4317&  1.5297&  1.6753&  1.2197&  1.2373& -17.6\\ \hline 
         0.6318&  1.5028&  1.6815&  1.1928&  1.2435& -50.7\\ \hline
    \end{tabular}

\end{table}

\section{Photoluminescence (PL) lineshape fitting}
The peak position and width were extracted from each PL spectra by fitting with a sum of Gaussian functions because of the large inhomogeneous broadening:
\begin{equation}
    I(E)=\sum_{j}\frac{A_j}{w_j\sqrt{\frac{\pi}{2}}}e^{-\frac{2(E-E_{cj})}{w_j^2}}
\end{equation}
where \textit{I} is intensity, \textit{E} is energy, \textit{E$_{cj}$} is peak position, \textit{w$_j$} is width and \textit{A$_j$} is the integrated area under the curve for \textit{j$^{th}$} peak. The fairly good fits obtained with such a lineshape, as shown in Fig. 2(a)-(c), indicates that this approximation is reasonable.
\begin{figure}[H]
  \includegraphics[width=15cm]{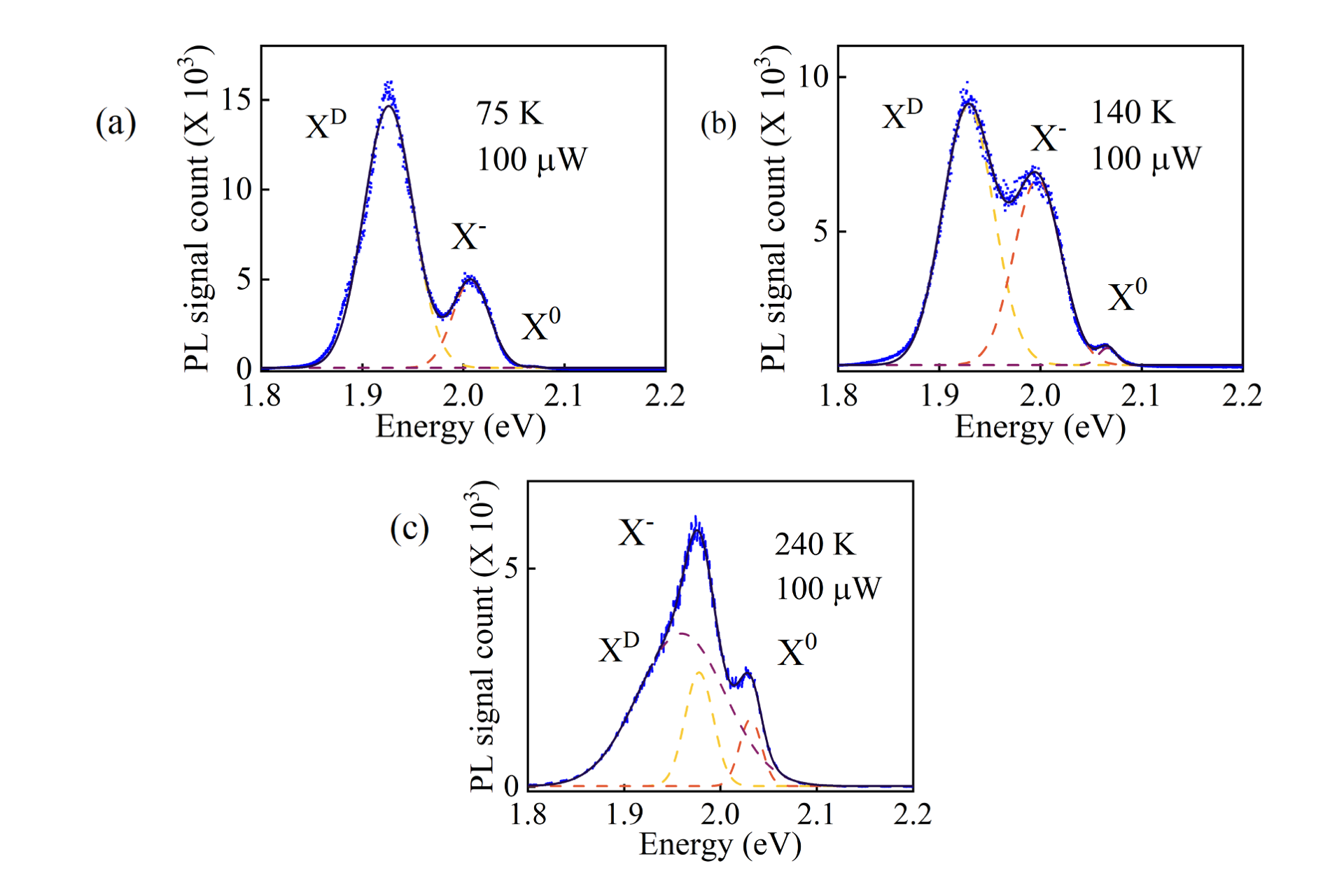}
  \centering
  \caption{(a)-(c) PL spectrum of ML WS$_2$ on C-99 substrate fitted with sum of Gaussians. The dashed lines represent the individual fitted components, solid black line is the cumulative fit and blue dots represent the raw data.
  \label{graph2}}
\end{figure}
\section{Power dependence of PL intensity}
\begin{figure}[H]
  \includegraphics[width=10cm]{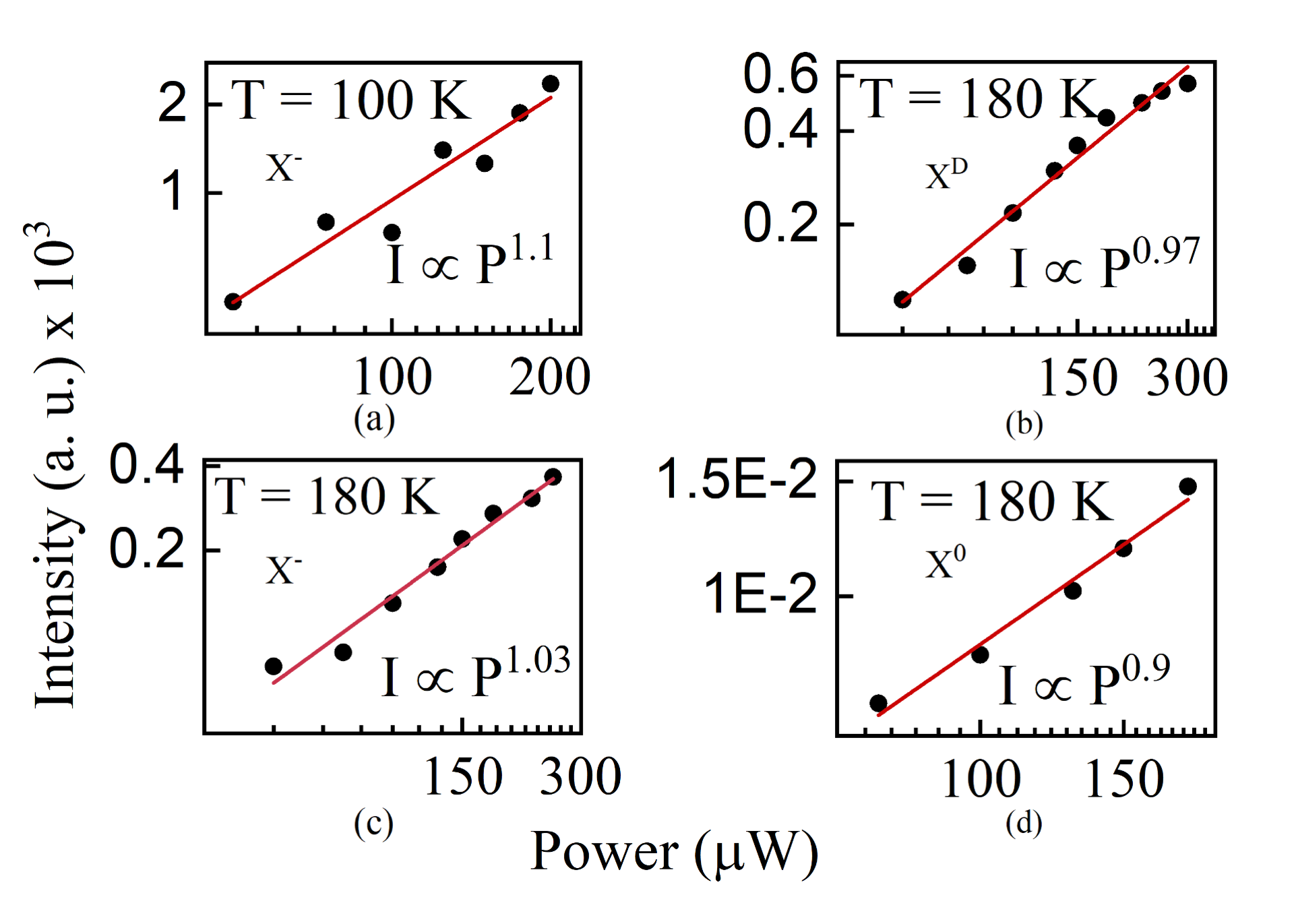}
  \centering
  \caption{Variation of integrated intensity (extracted by fitting each spectra Eq. 2) with increase in pump laser (514.5 nm) power for (a) X$^{-}$ at 100 K (b) X$^{D}$ at 180 K (c) X$^{-}$ at 180 K and (d) X$^{0}$ at 180 K. The data in (a)-(d) are fitted with the relation I $\propto$ P$^\gamma$ (red solid lines). The extracted exponent $\gamma$ is indicated in the respective plots.
  \label{graph3}}
\end{figure} 

\section{Power dependence of the peak position of X$^D$ peak}
\begin{figure}[H]
  \includegraphics[width= 6cm]{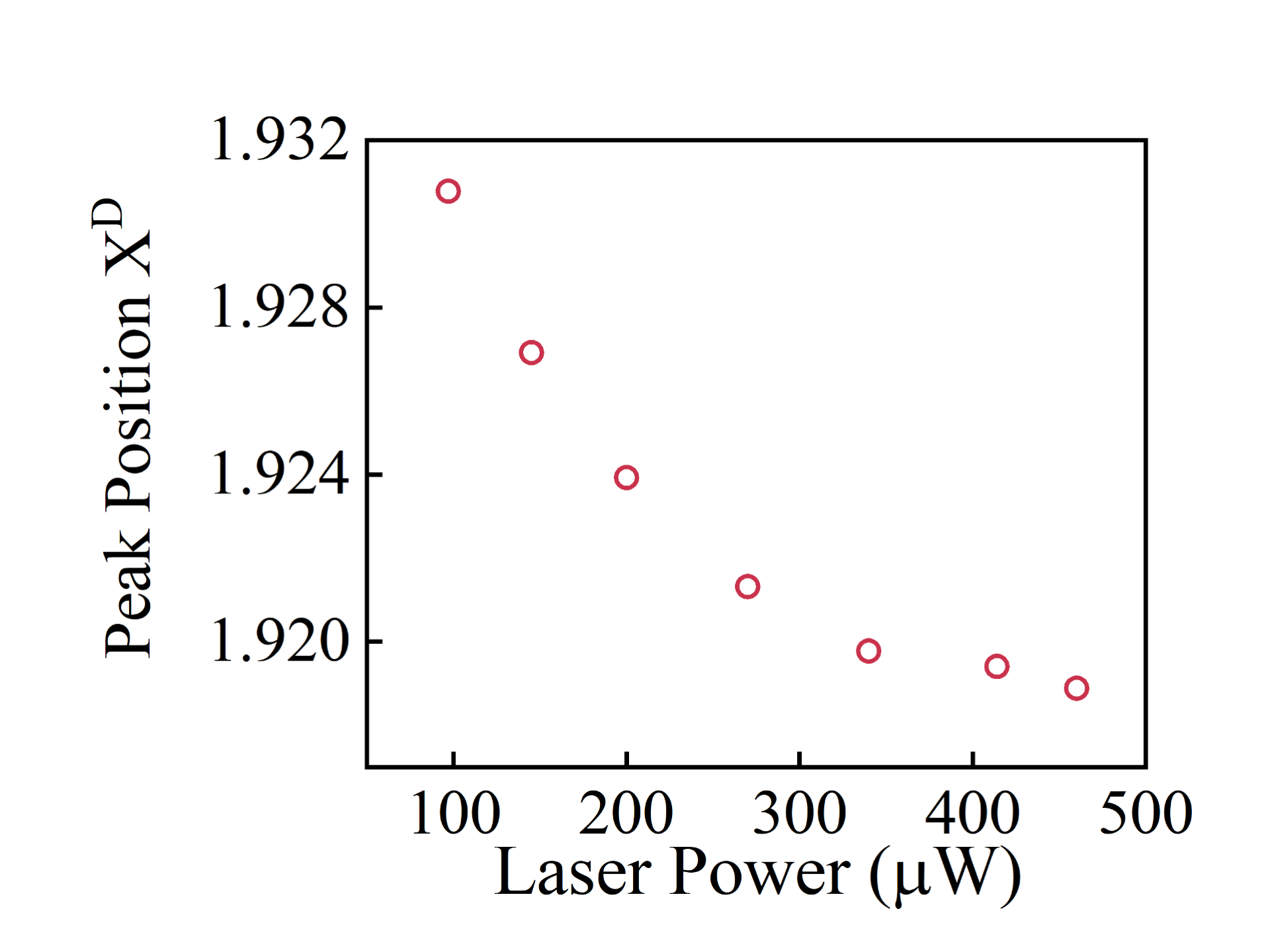}
  \centering
  \caption{Variation of emission peak energy of X$^D$ in ML WS$_2$ on C-99 substrate at 75 K with pump laser (514.5 nm) power.
  \label{graph4}}
\end{figure}
\section{ Raman lineshape fitting}
The various Raman peaks (E$^{'}$, 2LA, A$_{1}^{'}$) were analysed with multiple Lorentzian functions:
\begin{equation}
    I(E)=\sum_{j} \frac{2A_{j}w_{j}}{4\pi(E-E_{cj})^2+w_j^2}
\end{equation}
\begin{figure}[H]
   \includegraphics[width=15 cm]{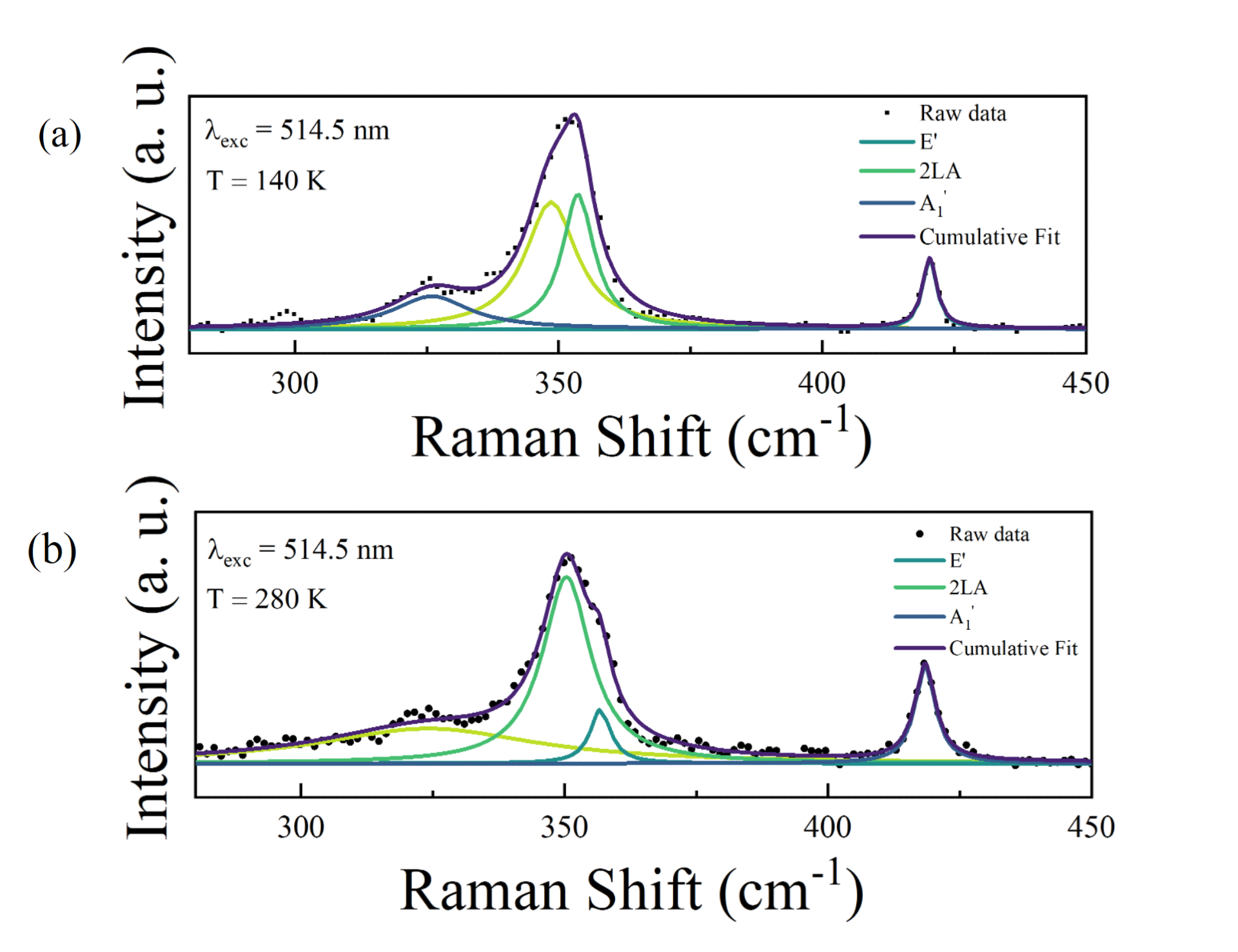}
   \centering
   \caption{(a)-(c) PL spectrum of ML WS$_2$ on C-99 substrate fitted with sum of Gaussians. The dashed lines represent the individual fitted components, solid black line is the cumulative fit and blue dots represent the raw data.
   \label{graph4}}
 \end{figure}
where \textit{I} is intensity, \textit{E} is energy, \textit{E$_{cj}$} is peak position, \textit{w$_j$} is width and \textit{A$_j$} is the integrated area under the curve for \textit{j$^{th}$} peak.
\section{Absence of Phonon Anomaly of the A$_1^{'}$ peak}
The redshift and increasing linewidth of Raman mode with temperature can be explained by anharmonic cubic equations\cite{joshi2016phonon,balkanski1983anharmonic}:
 \begin{equation}
    E_{ph}(T)= E_0 - C\Big(1+\frac{2}{e^\frac{E_0}{2k_BT}-1}\Big);
    \label{eq:4}
    \gamma_{ph}(T)= \gamma_0 + D\Big(1+\frac{2}{e^\frac{E_0}{2k_BT}-1}\Big)
    \label{eq:5}
\end{equation}
 \begin{figure}[H]
   \includegraphics[width=12 cm]{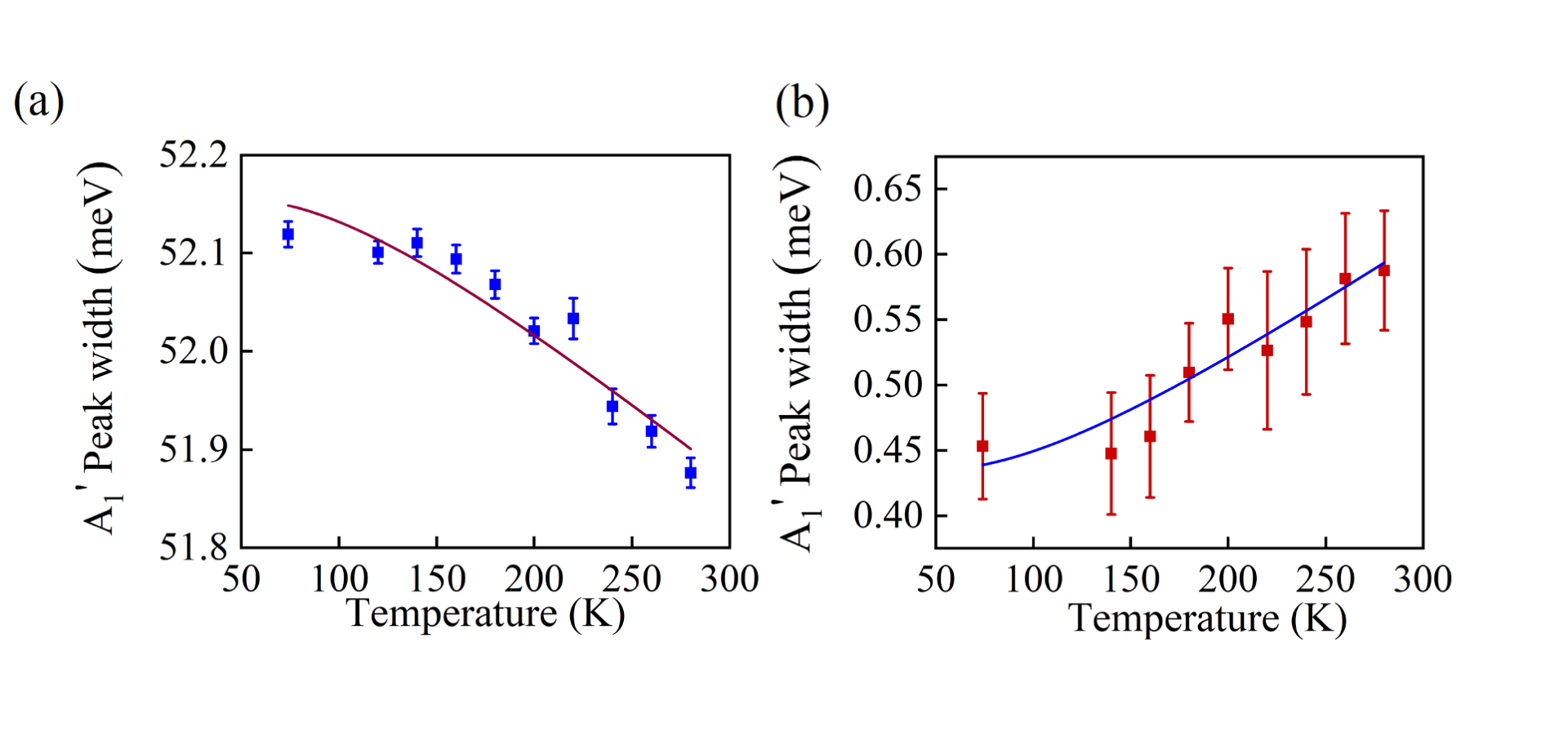}
   \centering
   \caption{Non-anomalous behaviour of A$^{'}_{1}$ peak 
(a) position and (b) width as a function of temperature. The solid lines in (a) and (b) are fitted lines with Eq 3 and 4 of main paper respectively.
   \label{graph4}}
 \end{figure}
\vspace{-1cm}

\section{Room temperature Raman map of WS$_2$ flake}
 \begin{figure}[H]
   \includegraphics[width=15 cm]{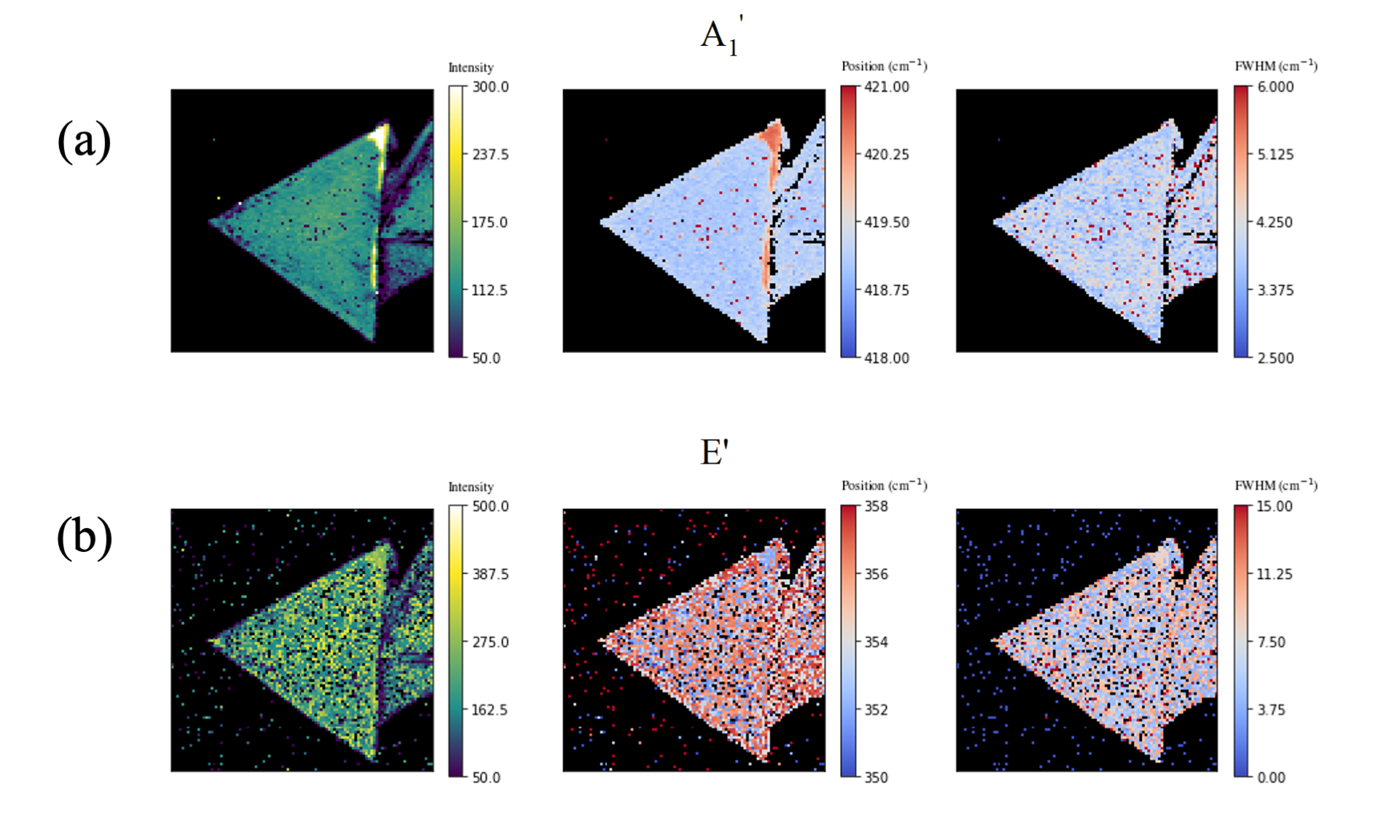}
   \centering
   \caption{ Room temperature Raman map of the ML WS$_2$ flake placed on top of C-99 substrate. Intensity, position and FWHM map of the (a) A$_1^{'}$ and (b) E$^{'}$ phonon mode.
   \label{graph5}}
 \end{figure}
  \begin{figure}[H]
   \includegraphics[width=10 cm]{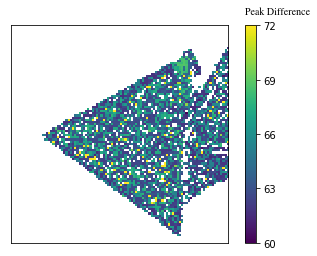}
   \centering
   \caption{ Raman peak position difference map of (A$_1^{'}$ -
   E$^{'}$) modes at room temperature.
   \label{graph5}}
 \end{figure}
 As the layer number increases, the A$_1^{'}$ mode blue shifts by a few wavenumbers and the E$^{'}$ mode redshift very gradually, as previously reported. Peak position difference of A$_1^{'}$ and E$^{'}$ in ML WS$_2$ is reported to be $\sim$ 65 cm$^{-1}$ while for bilayer and more the difference become $>$ 68 $cm^{-1}$\cite{doi:10.1021/nl3026357, chernikov2014exciton}. This shows that the WS$_2$ in our study is indeed monolayer. However, A$_1^{'}$ and E$^{'}$ mode redshifts due to tensile strain at a rate of $-$5.7 and $-$ 1.8 cm$^{-1}$/\%\cite{michail2023tuning}. The ML WS$_2$ on top of C-99 is under a tensile strain of 0.11\% which translates into a peak difference increase of 0.4 cm$^{-1}$. Thus effect of strain on the peak difference is negligible.    
\section{EXCITON-PHONON coupling}
The temperature dependence show a  characteristic redshift with decreasing temperature induced by a exciton-phonon coupling. This blue-shift can be described by a phenomenological model proposed by O'Donnell and Chen\cite{o1991temperature}:
\begin{equation}
    E(T)= E(0)-S\langle\hbar\omega\rangle\Big(\coth{\frac{\langle\hbar\omega\rangle}{k_{\mathrm{B}}T}}-1\Big)
    \label{eq:2}
\end{equation}
where E(T) is the resonance energy of X$^0$ at temperature T, `S' is dimensionless exciton-phonon coupling constant, $k_B$ is Boltzman constant and $\hbar\omega$ is the average phonon energy responsible for the coupling. By fitting the experimental data we obtained the parameters to be, E(0)= 2.0893 $\pm$ 0.0008 eV, S= 2.35 $\pm$ 0.09, $\langle\hbar\omega\rangle$= 29.45 $\pm$ 1.92 meV for C-48 and  E(0)= 2.0862 $\pm$ 0.003 eV, S= 1.87 $\pm$ 0.19, $\langle\hbar\omega\rangle$= 20.33 $\pm$ 6.01 meV for C-132.

From Figure 4d we know the value of strain in C-49 and C-132 is less than C-99. From  the DFT calculations in Figure 4c we know that for a strain value less than that of C-99, $\Delta$E $>$ 44 meV. The average phonon-energy $\langle\hbar\omega\rangle$ extracted from the above analysis for C-48 and C-99 is well below 44 meV. Therefore, we cannot observe X$^D$ in C-48 and C-132.
 \begin{figure}[H]
   \includegraphics[width=15 cm]{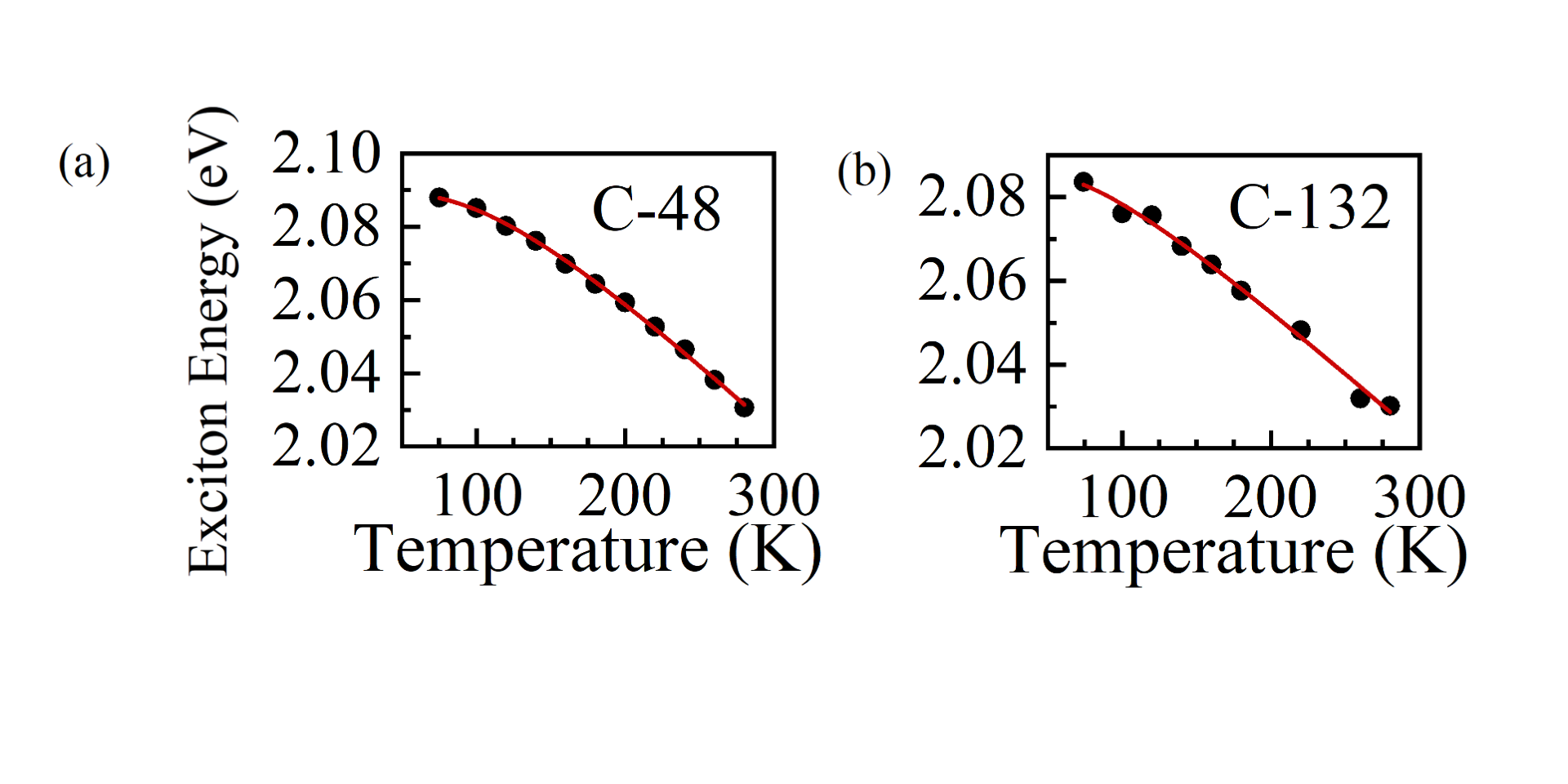}
   \centering
   \caption{Temperature dependence of the peak position of X$^0$ (black solid circles)  fitted with Eq. 5 (red line) in ML WS$_2$ placed on top of (a) C-48 and (b) C-132 substrate.
   \label{graph5}}
 \end{figure}
  \begin{figure}[H]
   \includegraphics[width=13 cm]{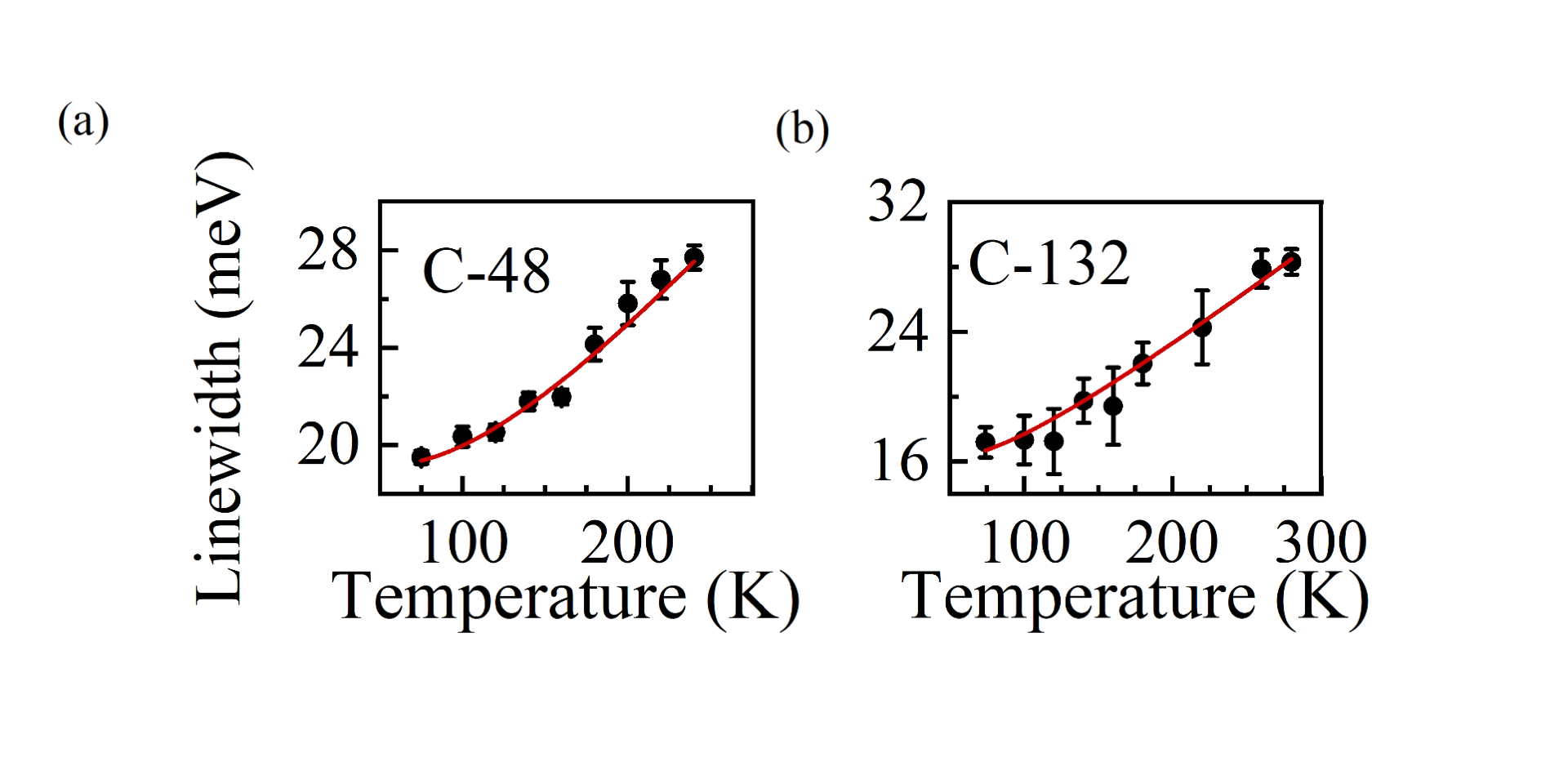}
   \centering
   \caption{Linewidth of X$^0$ as a function of temperature fitted with Eq. 6 in ML WS$_2$ placed on top of (a) C-48 and (b) C-132 substrate.
   \label{graph5}}
 \end{figure}
  To determine the strength of exciton-phonon coupling, the evolution of linewidth of X$^0$ was fitted by a phonon-induced broadening model\cite{selig2016excitonic,cadiz2017excitonic}:
\begin{equation}
    \gamma=\gamma_{0}+c_1T+\frac{c_2}{e^{\frac{\hbar\omega}{k_{B}T}}-1}
    \label{eq:3}
\end{equation}
where $\gamma_0$ is the broadening at T= 0 K, the linear term in T is due to the interaction of acoustic phonon modes (LA and TA) and the last term is the interaction term with optical phonon mode E$^{'}$. The linear term is neglected compared to the last term which is proportional to the the Bose function for optical phonon mode occupation\cite{arora2015excitonic}. c$_2$ is the measure of the exciton-photon coupling strength. $\hbar\omega$ obtained by fitting Eq. 5  was used for fitting Eq. 6. 
 \section{PL map of ML WS$_2$ flake on top of C-99 substrate}
  \begin{figure}[H]
   \includegraphics[width=15 cm]{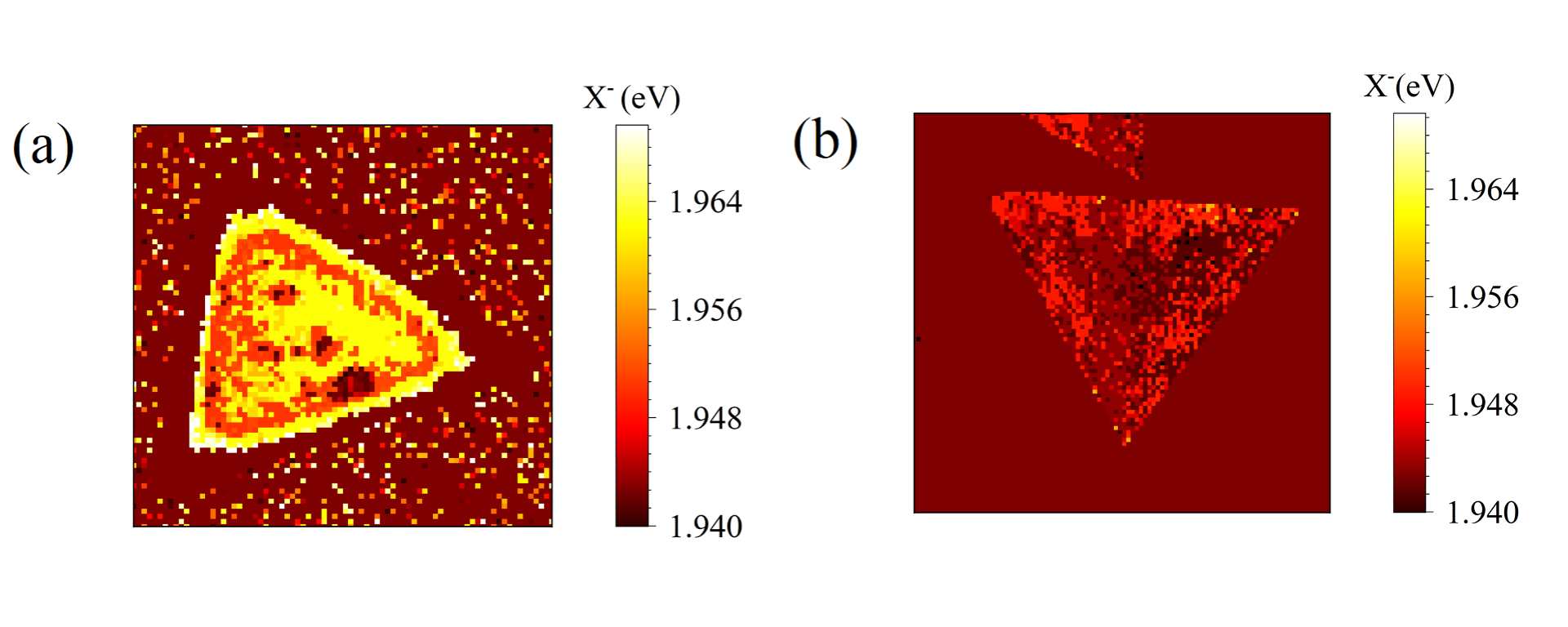}
   \centering
   \caption{ Trion (X$^-$) position map in ML WS$_2$ placed on top of (a) Si/SiO$_2$ and (b) C-99 substrate.
   \label{graph5}}
 \end{figure}
 \begin{figure}[H]
   \includegraphics[width=15 cm]{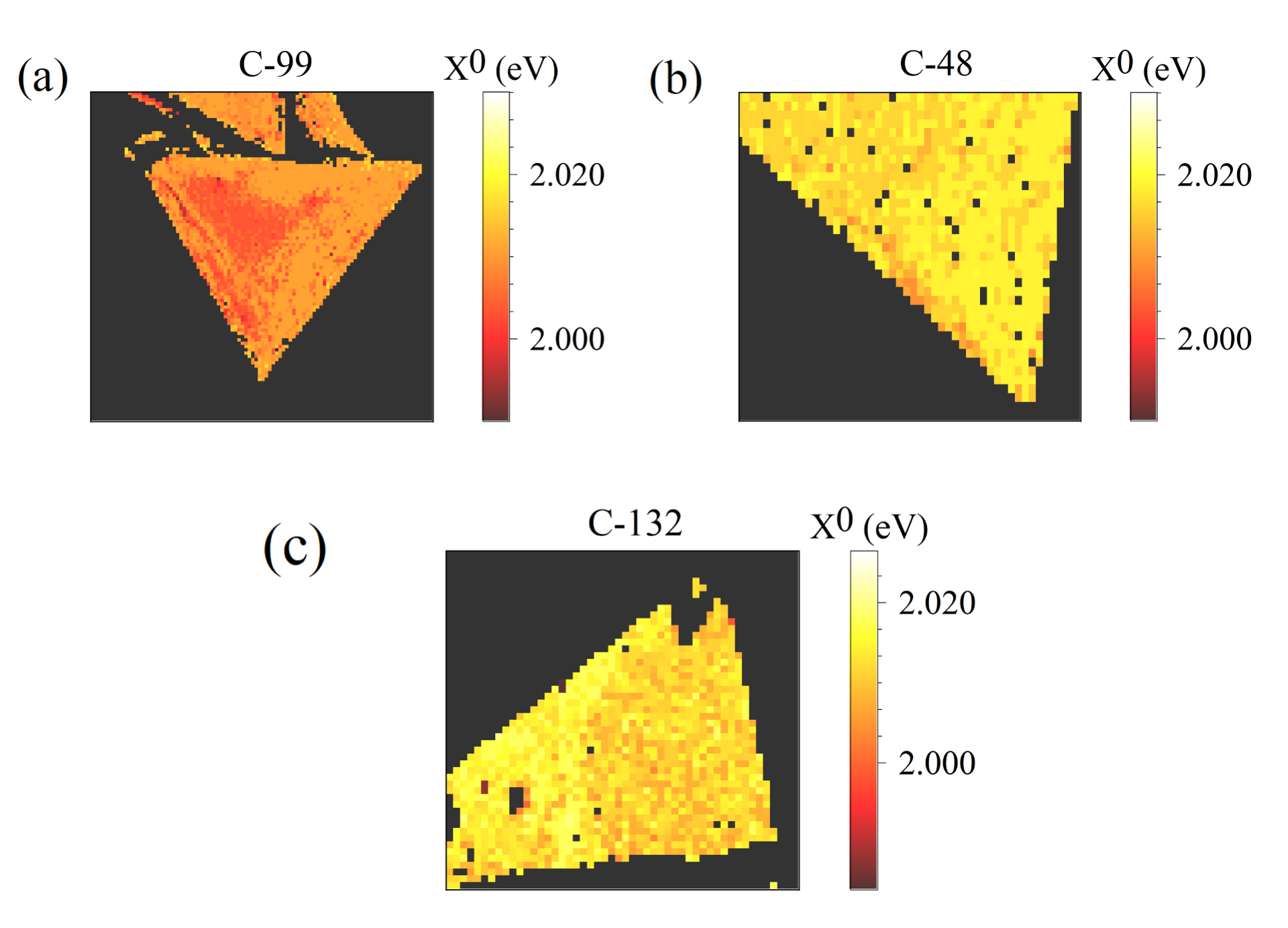}
   \centering
   \caption{ Exciton (X$^0$) position map in ML WS$_2$ placed on top of (a) C-99 (b) C-48 and (c) C-132 substrate.
   \label{graph5}}
 \end{figure}

 \section{Exciton binding energy calculation}
 The exciton binding energy was calculated by 
determining the quasiparticle band gap corresponding to the
energy of a separated electron-hole pair. This is 
done in semiconductors by fitting the excitonic
peaks to a hydrogenic Rydberg series\cite{klingshirn2012semiconductor}. In 2D, this
hydrogen model employs an effective mass Hamiltonian\cite{chernikov2014exciton}:

\begin{equation}
    H= -\hbar^2\frac{\nabla^2_r}{2\mu}+V_{eh}(r)
\end{equation}
where $\frac{1}{\mu}\!=\!\frac{1}{m_e}+\frac{1}{m_h}$ is reduced mass of exciton, m$_h$ and m$_e$ are the effective masses of hole and electron respectively. and $V_{eh}(r)=\frac{-e^2}{\epsilon r}$ is a locally screened attractive electron-hole interaction. This model predicts exciton transition energies of E$_g$-E$_b^{(n)}$, where E$_g$ is the quasiparticle gap and 
\begin{equation}
    E_b^{(n)}= \frac{\mu e^4}{2\hbar^2\epsilon^2(n-\frac{1}{2})^2}
\label{eq:7}
\end{equation}

 \bibliographystyle{apsrev4-2}
\bibliography{refs}